\documentclass[aps,prb,reprint,superscriptaddress,longbibliography,showpacs,nobibnotes,nofootinbib]{revtex4-2}

\usepackage[utf8]{inputenc}
\usepackage[T1]{fontenc}
\usepackage[]{lmodern}
\usepackage[]{microtype}

\usepackage[english]{babel}
\usepackage[]{csquotes}

\makeatletter
\adddialect\l@en\l@english
\makeatother

\usepackage[]{amsmath}
\usepackage[]{amssymb,bbold}
\usepackage[mathscr]{euscript}

\usepackage{siunitx}
\usepackage[]{graphicx}
\usepackage[]{tabularx}
\usepackage{tikz-cd}
\usepackage[export]{adjustbox}
\usepackage[figurename=Figure]{caption}
\usepackage[]{subcaption}

\usepackage{ragged2e}
\DeclareCaptionJustification{plain}{\justifying}
\usepackage[]{hyperref}
\hypersetup{
    colorlinks=true,
    urlcolor=blue,
    citecolor=blue,
    linkcolor=blue
}
\usepackage[capitalize]{cleveref}
\crefname{table}{Tab.}{Tab.}

\usepackage{xcolor}
\definecolor{nodetuning}{rgb}{0.403, 0.678, 0.325}
\definecolor{interactdetuning}{rgb}{0.824, 0.208, 0.169}
\definecolor{localdetuning}{rgb}{0.949, 0.663, 0.231}
\definecolor{fillingdetuning}{rgb}{0, 0, 0.961}
\definecolor{greedydetuning}{rgb}{0.459, 0.078, 0.4902}

\DeclareMathOperator*{\argmax}{argmax} 
\DeclareMathOperator{\sgn}{sgn}
\DeclareMathOperator{\Tr}{Tr}
\DeclareMathOperator{\GE}{\mathsf{GE}}
\DeclareMathOperator{\DC}{\mathsf{DC}}
\DeclareMathOperator{\LC}{\mathsf{LC}}
\DeclareMathOperator{\GEC}{\mathsf{GEC}}

\newcommand{\ket}[1]{\ensuremath{{\mkern -3mu}\left\lvert#1\right\rangle}{\mkern -3mu}}
\newcommand{\bra}[1]{\ensuremath{{\mkern -3mu}\left\langle#1\right\rvert{\mkern -3mu}}}
\newcommand{\mel}[3]{\ensuremath{\bra{#1}#2\ket{#3}}}

\newcommand{\changes}[1]{{\color{blue}#1}}
\renewcommand{\changes}[1]{#1}

\begin{document}
	
\title{Graph theory and tunable slow dynamics in quantum East Hamiltonians}

\author{Heiko Georg Menzler}
\affiliation{Institut f\"{u}r Theoretische Physik, Georg-August-Universit\"{a}t G\"{o}ttingen, D-37077 G\"{o}ttingen, Germany}

\author{Mari Carmen Ba\~nuls}
\affiliation{Max-Planck-Institut f\"{u}r Quantenoptik, D-85748 Garching, Germany}
\affiliation{Munich Center for Quantum Science and Technology (MCQST), Schellingstrasse 4, D-80799 München, Germany}

\author{Fabian Heidrich-Meisner}
\affiliation{Institut f\"{u}r Theoretische Physik, Georg-August-Universit\"{a}t G\"{o}ttingen, D-37077 G\"{o}ttingen, Germany}

\begin{abstract}
    We show how graph theory concepts can provide an insight into the origin of slow dynamics in systems with kinetic constraints.
    In particular, we observe that slow dynamics is related to the presence of strong hierarchies between nodes on the Fock-space graph in the particle occupation basis, which encodes configurations connected by a given Hamiltonian.
    To quantify hierarchical structures, we develop a measure of centrality of the nodes, which is applicable to generic Hamiltonian matrices and inspired by established centrality measures from graph theory.
    We illustrate these ideas in the quantum East (QE) model.
    We introduce several ways of detuning nodes in the corresponding graph that alter the hierarchical structure, defining a family of QE models.
    We numerically demonstrate how these detunings affect the degree of non-ergodicity on finite systems, as evidenced by both the time dependence of density autocorrelations and eigenstate properties in the detuned QE models.
    % keywords: slow dynamics, ergodicity, graph theory, centrality measures
\end{abstract}

\date{\today}

\maketitle

\section{Introduction}
Generic interacting closed quantum systems are expected to relax to thermal equilibrium~\cite{DAlessio2016}, consistent with the \emph{eigenstate thermalization hypothesis} (ETH)~\cite{Deutsch1991,Srednicki1994,Srednicki1999,Deutch2018,Rigol2008}.
Violations of ETH lead to non-ergodic dynamics, where the system does not thermalize and information of an initial configuration may be partly preserved even in local measurements~\cite{Kaufman2016}. Often, this manifests itself in  autocorrelation functions not decaying to zero
as a function of time.

For a given Hamiltonian, non-ergodic behavior can be detected from eigenstate properties, such as matrix elements of local observables, statistics of spectral correlations of the eigenenergies, and entanglement in eigenstates~\cite{DAlessio2016}.
The best studied candidates for fully non-ergodic systems are many-body localized systems~\cite{Nandkishore2015,Abanin2019,Sierant2024,Suntajs2020,Abanin2021,Sels2021,DeRoeck2017,Morningstar2022,deRoeck2024}. More recently, the interest has shifted to models with constrained dynamics that can lead to weak or strong ergodicity breaking~\cite{Turner2018,Sala2020, Khemani2020,Moudgalya2022}.
In these systems, a subset of all eigenstates violates ETH, while the number of such states can be intensive or extensively large.

Constrained dynamics can be caused by kinetic constraints in the Hamiltonian, as realized by, e.g., Rydberg blockade~\cite{Valado2016,Magoni2021}, dipole conservation~\cite{Sala2020,Khemani2020}, gauge constraints~\cite{Karpov2021,Brenes2018,Jeyaretnam2024}, or constraints inspired from classical physics~\cite{VanHorssen2015,Lan2018,Garrahan2018}.
Examples of consequences of constrained dynamics are quantum scars~\cite{Turner2018} and Hilbert-space fragmentation~\cite{DeTomasi2019,Sala2020,Khemani2020}.
In the former, one or typically countably many eigenstates exhibit sub-volume law scaling of entanglement.
In the latter, the Hilbert space breaks up into exponentially many subspaces in specific basis sets.
Both cases have recently been accessed in quantum-simulator experiments~\cite{Bernien2017,Scherg2021,Kohlert2023,Su2023,Adler2024,Kim2024,Wang2025,Honda2025}.

In our work, we investigate slow dynamics in systems with kinetic constraints that have an irreducible subspace, i.e., no Hilbert-space fragmentation, which frequently occurs in kinetically constrained models (KCMs), see, e.g.,~\cite{Wang2023,Brighi2023}.
In the 
case of classical KCMs, there is a natural connection to graph theory since the dynamics is only sensitive to the allowed transitions between the classical configurations, which can thus be expressed as an underlying graph~\cite{Ritort2003,Garrahan2011}.
In this work, we further exploit the connection to graph theory and extend it to quantum models, motivated also by earlier work (see, e.g.,~\cite{Roy2020,Desaules2022}).

Given a quantum Hamiltonian and a many-body basis, we can associate it with a Fock-space graph. This graph alone can in general
not fully determine the dynamical behavior under the unitary time evolution, 
since it only captures off-diagonal Hamiltonian terms.
For the classes of Hamiltonians studied in this work, it is however the competition between diagonal and off-diagonal terms that establishes a hierarchy between graph nodes correlated to the degree of slow dynamics.

Here, we account for diagonal terms 
by utilizing graph-theoretical tools.
 We introduce a specific hierarchy measure, the graph energy centrality ($\GEC$), inspired by the Laplacian centrality from graph theory~\cite{Qi2012}, that quantifies the importance of individual basis configurations.
Via numerical simulations, we establish a connection between the spread of graph energy centrality and the degree of non-ergodicity, as measured by spectral properties, eigenstate entanglement, and the time-dependence of autocorrelation functions.

In order to illustrate these ideas, we consider the quantum East model, for which the existence of ergodic regions and regions with slow dynamics is well established~\cite{VanHorssen2015,Pancotti2020}, at least on finite systems.
The Hamiltonian of the QE model reads
\begin{align}
    \label{eq:hamiltonian_qeast_original}
    H_\mathrm{QE}
    =
    - \frac{1}{2} \sum\limits_{\ell = 1}^{L-1} n_\ell (e^{-s}\sigma^x_{\ell+1} - \openone)\,,
\end{align}
where $e^{-s}/2$ is the transition amplitude, with $s$ real, and $n_\ell$ is the projector onto states with one particle at lattice site $\ell$, or, equivalently, the density operator.
The operator $\sigma^x_\ell$ is the $x$-Pauli matrix acting on site $\ell$, which adds or removes one particle on that site.
In order to obtain an irreducible subspace, we impose
so-called East boundary conditions, i.e., there is always a particle on site $\ell=0$ [details are given in \cref{sec:introduction_model} and Eq.~\eqref{eq:HQE_actual}].

\begin{figure}[t]
    \centering
    \includegraphics[width=0.95\linewidth]{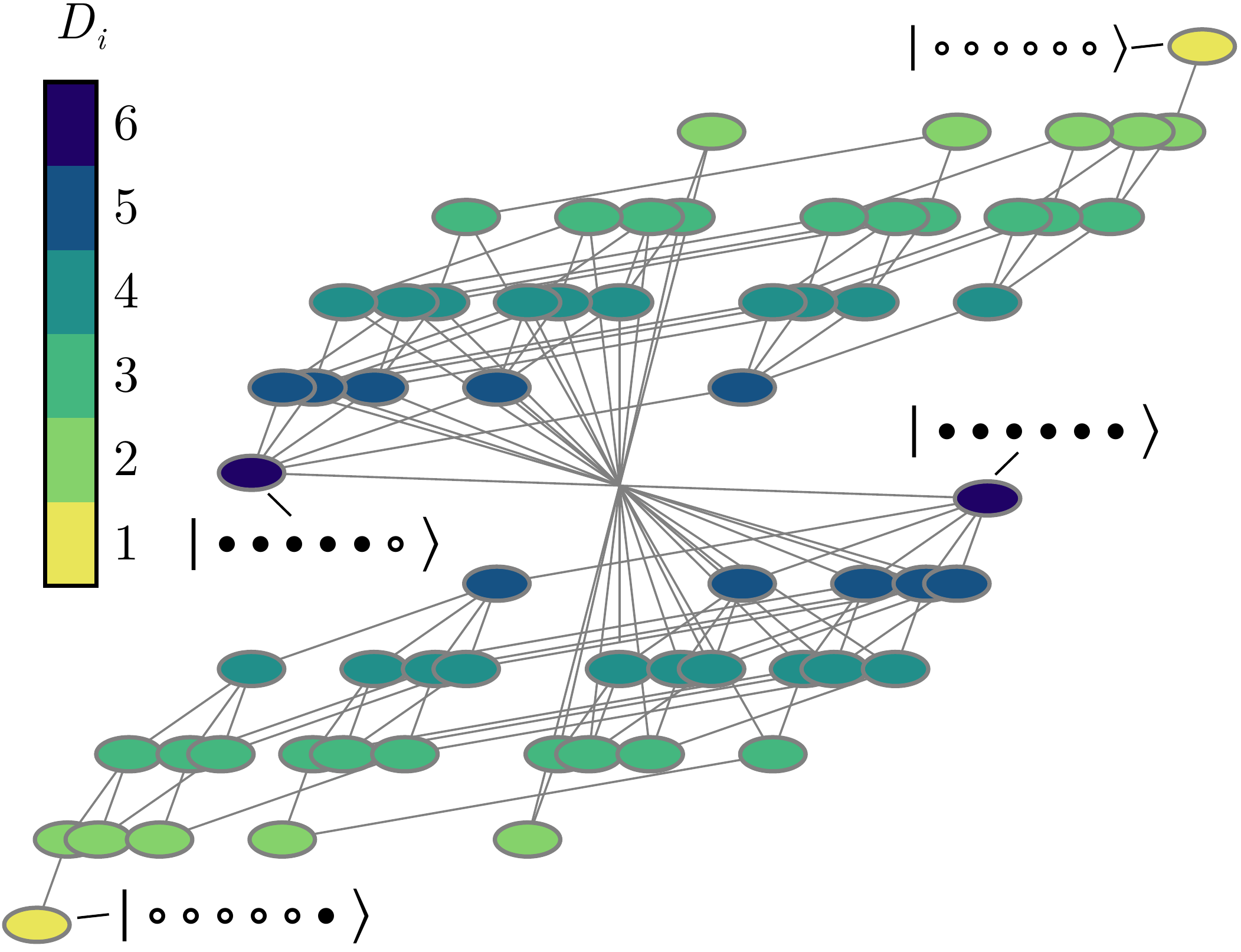}
    {\phantomsubcaption\label{fig:qe_graph_degree}}
    \vspace{1em}
    \caption{
        Fock-space graph structure of the QE model for $L=6$ with East boundary conditions [see \cref{eq:HQE_actual}]
        set up in the joint eigenbasis of all $n_\ell$ operators.
        Darker colors indicate a higher degree of connectivity $D_i$ of a given node.
        In the chosen presentation, the graph is (point-)symmetrical along the horizontal axis and the states in the two subgraphs differ only on the last (real-space) lattice site. Corresponding nodes on the upper and lower subgraph differ in the particle number by one, but they do have the same number of particles that allow for transitions on right neighbors in the real-space lattice. 
        Therefore, corresponding nodes have the same degree of connectivity on both subgraphs. On the last site, the presence of a particle does not lead to additional transitions for the chosen boundary conditions.
    }
    \label{fig:qe_graph}
\end{figure}

We set up the Fock-space graph in the particle occupation basis with states $\ket{i}$, i.e., the joint eigenbasis of all $n_\ell$ operators.
An edge is drawn between two nodes if they are connected by an off-diagonal term of Eq.~\eqref{eq:hamiltonian_qeast_original}. The resulting graph is illustrated in Fig.~\ref{fig:qe_graph} (see~\cite{Roy2020}).
In this model, the degree of connectivity $D_i$ of a node, i.e., the number of configurations directly connected to $\ket{i}$ by the Hamiltonian \cref{eq:hamiltonian_qeast_original},  is essentially determined by the number of particles
(see the caption of Fig.~\ref{fig:qe_graph} and Sec.~\ref{sec:introduction_model} for details).
In the QE model itself, the slow dynamics is 
a result of the competition between the off-diagonal and the diagonal term. This competition can be viewed as a 
(diagonal) detuning of nodes with different particle numbers with respect to each other.

We introduce a family of quantum East models 
by adding different detuning terms to the standard QE model. In each of them, nodes are detuned by an operator $P$, diagonal in the occupation eigenbasis
\begin{align}
H_{\mathrm{QE},P} = H_\mathrm{QE} + z P \,,
\label{eq:havenoname}
\end{align}
where $z$ is the detuning strength. We set $z$ to unity throughout this work.

Our work has three main results. First, 
we show that the diagonal detuning of basis states is able to induce or suppress slow dynamics in finite-size systems.
Second, 
by analyzing four different choices of $P$ at fixed value $z=1$, we observe that 
the degree of non-ergodicity correlates with the standard deviation of graph energy centrality.
Third, we show that graph energy centrality can be computed efficiently on much larger system sizes than what is accessible in exact diagonalization, different from most other eigenstate- and eigenspectrum-based measures.

The rest of this exposition is organized as follows.
In \cref{sec:introduction_model}, we present the quantum East model and summarize known results.
In \cref{sec:qegraph}, we discuss the connection of the QE model to graph theory and introduce centrality measures.
There, we propose graph energy centrality as a key concept in our work.
In \cref{sec:graph_detuning}, we define several detuning protocols that allow us to tune the dynamics from entirely ergodic to strongly non-ergodic (on finite systems).
We supplement this with an analysis of density autocorrelations.
\Cref{sec:slow_dynamics_entanglement} 
provides a discussion of how eigenstate properties change between the different detuning protocols.
Finally, we summarize our results and point out future research directions in \cref{sec:summary}.
\cref{sec:appendix_construct_many_body_graphs} describes the construction of graphs for many-body systems.
\cref{sec:appendix_gec_pdf} provides details on the calculation of graph energy centrality for a number of quantum East-like models.
In \cref{sec:appendix_filling} we discuss details of detuning protocols.

\section{Quantum East Model and Observables}
\label{sec:introduction_model}

\begin{figure}[t]
    \centering
    \includegraphics[width=0.98\linewidth]{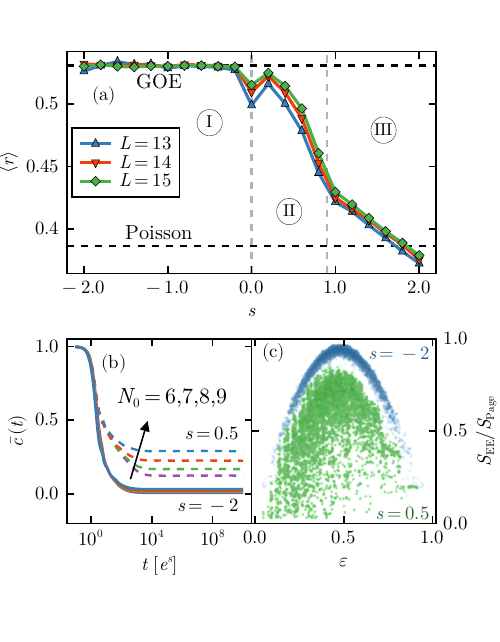}
    {\phantomsubcaption\label{fig:intro_figures_gapratio}}
    {\phantomsubcaption\label{fig:intro_figures_autocors}}
    {\phantomsubcaption\label{fig:intro_figures_EEs}}
\vspace{-1.5cm}
    \caption{
        Illustration of constrained dynamics in the quantum East model.
        \subref{fig:intro_figures_gapratio}~Gap ratio $\langle r \rangle$ as a function of $s$ averaged over 50 realizations of an ensemble of quantum East Hamiltonians with small variations in the chemical potential $\mu = 1 + \delta \mu$ with $\delta \mu$ sampled from $\delta \mu \in [-0.05, 0.05]$, for $L=13,14,15$, and averaged over all eigenstates.
        There are three regimes:
        (I) an ergodic one for $s<0$, (II) a region with slow dynamics for $0<s\lesssim 1$, and (III) the large-$\mu$ regime for $s \gtrsim 1$.
        In this work, we are interested in the former two regions.
        The horizontal dashed lines indicate the values of the gap ratio for the GOE and the Poisson distribution, respectively~\cite{Atas2013}.
        \subref{fig:intro_figures_autocors}~Density autocorrelation function $\overline{c}(t)$ averaged over initial product states with an initial particle number of $N_0$ for $L=12$ in the ergodic regime ($s=-2$, solid lines) and the slow-dynamics regime ($s=0.5$, dashed lines)~\cite{VanHorssen2015}.
        For $s=0.5$, $\overline{c}(t)$ converges to a non-zero value that depends strongly on $N_0$, whereas for $s=-2$ it decays quickly.
        \subref{fig:intro_figures_EEs}~Half-chain entanglement entropy $S_\mathrm{EE}$ normalized by the Page value $S_\mathrm{Page} = \frac{L\log( 2 ) - 1}{2}$ as a function of energy density $\varepsilon$ for $s=-2$ (blue) and $s=0.5$ (green) for $L=12$.
        In contrast to the $s=-2$ case, the case with $s=0.5$ hosts many eigenstates with atypically low entanglement~\cite{Pancotti2020}.
    }
    \label{fig:intro_figures}
\end{figure}

The classical East model~\cite{Jaeckle1991,Munos1998,Garrahan2002,Ritort2003} has been long investigated as a prototype of kinetically constrained spin glasses.
Its quantum version, the QE model, was introduced in Ref.~\cite{VanHorssen2015} and 
has gained attention in the recent literature due the simplicity of its constraint and the existence of slow dynamics in the complete absence of conservation laws or symmetries~\cite{VanHorssen2015,Pancotti2020}.
Recently, there has also been interest in models derived from or similar to the QE model~\cite{Roy2020,Bertini2023,Geissler2023,Wang2023,Brighi2023,Causer2025,Brighi2024,Bertini2024,Ganguli2024,Aditya2025} and in Floquet quantum East models~\cite{Bertini2023,Bertini2024}.
For the aforementioned reasons, the QE model has become an important example for slow dynamics in the literature of quantum systems with kinetic constraints.

In the Hamiltonian from Eq.~\eqref{eq:hamiltonian_qeast_original}, the projector $n_\ell$ plays the role of the constraint: 
a particle can only be added to or removed from the system if its left-neighboring site is occupied.
Since in the absence of a particle to the left, no dynamics occurs at all, one also refers to this as a facilitating constraint.
The allowed transitions in the QE model can be visualized as 
\begin{align}
    \ket{\dots \bullet \circ \dots} \leftrightharpoons \ket{\dots \bullet \bullet \dots}
\,,
\end{align}
where $\bullet$ indicates that there is a particle, while $\circ$ indicates 
an empty site.

While the Hamiltonian in \cref{eq:hamiltonian_qeast_original} is the original form (without boundary conditions) as it was introduced in Ref.~\cite{VanHorssen2015}, it is instructive to introduce a prefactor $\mu$ for the diagonal term, resulting in the following modified version
\begin{align}
    \label{eq:hamiltonian_qeast}
    H_{\mathrm{QE},\mu}
    =
    - \frac{e^{-s}}{2} \sum\limits_{\ell = 1}^{L-1}
        n_\ell \sigma^x_{\ell+1} 
    +
    \frac{\mu}{2} \sum\limits_{\ell = 1}^{L-1} n_\ell\,.
\end{align}
The parameter $\mu$ can be understood as a (dimensionless) chemical potential controlling the number of particles in the system, which, except for an intensive boundary effect at the last lattice site $L$, is the number of satisfied constraints.
In the case when $\mu = 1$, Eq.~\eqref{eq:hamiltonian_qeast} reduces to Eq.~\eqref{eq:hamiltonian_qeast_original}.

For open boundary conditions, the Hilbert space splits into $L+1$ subspaces identified by the location of the first particle, with all empty sites to its left being non-dynamical.
Additionally, the Hamiltonian commutes with $\sigma^x_L$, which splits each such sector into two.

To focus on a single sector for $L$ dynamical sites, we 
consider \emph{East} boundary conditions (EBC), corresponding to the addition of a non-dynamical, zeroth lattice site occupied by a particle, which facilitates the dynamics on the first site $\ell=1$, and an equally non-dynamical ($L+1$)-th site in a $\sigma^x_{L+1}$ eigenstate, with eigenvalue $\gamma=\pm 1$, which lifts the remaining symmetry.
Thus, we obtain the following Hamiltonian~\cite{Pancotti2020}
\begin{align}
H_\mathrm{QE}^\mathrm{EBC}= H_{\mathrm{QE}, \mu} - \frac{1}{2}\left(
     e^{-s}\sigma^x_1- \mu\openone
\right)
- 
\frac{n_L}{2}\left(
   \gamma e^{-s} - \mu
\right).
\label{eq:HQE_actual}
\end{align}
In our numerical calculations, we  implement $H_\mathrm{QE}^\mathrm{EBC}$ using the positive sign ${\gamma = +1}$.
With East boundary conditions, the quantum East model has 
a $2^L$-dimensional irreducible Hilbert space, i.e., which cannot be divided into further symmetry blocks.
Whenever $H_{\mathrm{QE},\mu}$ is used in our simulations, then we employ EBC as in Eq.~\eqref{eq:HQE_actual}.

The model with $\mu=1$ exhibits a quantum phase transition at $s=0$~\cite{VanHorssen2015,Pancotti2020}.
Regarding its nonequilibrium dynamics, the QE model with $\mu=1$ exhibits three regimes on finite system sizes:
(i) an ergodic regime for $s<0$, (ii) a regime with slow dynamics for $0<s\lesssim 1$, and (iii) a quasi-localized regime for $s\gtrsim1$~\cite{Pancotti2020,Menzler2024}.
We illustrate these known results by showing the gap ratio, the 
time dependence of density autocorrelations and the eigenstate entanglement in \cref{fig:intro_figures}.

As a standard measure of non-ergodicity, the gap ratio is defined as~\cite{Oganesyan2007,Atas2013}
\begin{align}
r_n = \frac{\min(\delta_{n-1}, \delta_n)}{\max(\delta_{n-1}, \delta_{n})}\,, \quad {\delta_n = E_{n} - E_{n-1}}\,,
\end{align}
where $E_n$ are the eigenvalues of the Hamiltonian in ascending order.
In order to obtain data with small fluctuations, it is useful to consider the extension~\eqref{eq:hamiltonian_qeast} of the QE Hamiltonian.
We average $r_n$ over an ensemble of 
systems
with different values of the chemical potential $\mu=1 + \delta \mu$
with ${\delta \mu \in [-0.05, 0.05]}$.
Drawing 50 samples from this ensemble and averaging $r_n$ over the whole spectrum for every sample, we 
obtain $\langle r\rangle $.

The results are shown in \cref{fig:intro_figures_gapratio}.
For $s<0$, the gap ratio follows the prediction of the Gaussian orthogonal ensemble (GOE), while for $s>0$, the value of $\langle r \rangle$ decreases, ultimately below the value expected for a Poisson distribution.
The small dip of $\langle r \rangle $ at $s=0$ is due to an additional symmetry at this point that is not lifted in our simulations~\cite{Pancotti2020}.

When the diagonal term dominates, i.e., for $s\gg1 $, the spectrum splits up into bands separated by gaps of order $\mu/2$.
The crossover into this regime is seen as a kink in the average gap ratio at $s\approx 1$ and has also been observed from other quantum-chaos measures in~\cite{Menzler2024}.

The separation of these bands is an intensive scale and
therefore, 
in the thermodynamic limit, a continuous spectrum will result
and the persistence of this band structure is not expected (see also~\cite{DeTomasi2019}).
Therefore, we do not further discuss the regime of $s\gtrsim 1$, but focus on the comparison of the ergodic regime ($s < 0$) and the slow-dynamics regime ($0 < s \lesssim 1$).

The presence of the ergodic and the slow-dynamics regimes manifests itself in density autocorrelations.
We define the density autocorrelation function as 
\begin{align}
\label{eq:autocorrelation_function}
    c(t) = \frac{1}{N_0}\sum_{\ell=1}^{L} 2 \langle n_\ell(t) n_\ell(0) \rangle - 1
\,,
\end{align}
where $N_0 = \sum_{\ell=1}^{L} \langle n_\ell(0) \rangle$.%
\footnote{
  We exclude the state with $N_0=0$.
}
To remove 
short-scale fluctuations in the correlations and thus extract information about the long-time behavior of $c(t)$, we compute the time average $\overline{c}(t) = 1/t \int_0^t \mathrm{d}t^\prime\, c(t^\prime)$.
For $s < 0$, the system exhibits fast dynamics indicated by quickly decaying autocorrelation functions, while for $s > 0$ the system shows slow dynamics, characterized by non-decaying autocorrelations $\overline{c}(t)$ on finite systems, see \cref{fig:intro_figures_autocors}~\cite{Pancotti2020}.

With regard to the long-time limit of $c(t)$ computed in individual initial states, note that even in the ergodic regime, non-zero and \emph{negative} long-time values can occur.
This is a consequence of both finite system size and lack of particle-number conservation in our system, and therefore $c(t) \to 0$ is only ensured
after averaging over initial states. Nonetheless, this is a quantitatively minor effect compared to the robust and large values of $c(t)$ in the slow-dynamics regime $0<s\lesssim 1$.
In the ergodic regime, the spread of nonzero long-time values of $c(t)$ is expected to scale fast to zero  as the system size increases, whereas non-ergodic states should exhibit a slower or no decay with $L$.
These details are not quantitatively relevant for the results
discussed in our work.

Finally, we turn to the entanglement entropy computed in eigenstates.
We consider the half-chain entanglement entropy defined as
\begin{align}
\label{eq:entanglement_entropy}
    S_\mathrm{EE} = -\Tr(\rho_A \log \rho_A)\,,
\end{align}
where $\rho_A = \Tr_{\bar A}(\ket{\psi} \bra{\psi})$ is the reduced density matrix of part $A$, chosen to be one half of the chain, and $\ket{\psi}$ is an eigenstate of $H$.
We rescale $S_\mathrm{EE}$ using the Page value~\cite{Page1993} for the half-chain subsystem $S_\mathrm{Page} = \frac{L\log(2) - 1}{2} $.

In \cref{fig:intro_figures_EEs}, we show exemplary results for $S_\mathrm{EE}$ as a function of energy density $\varepsilon$ for $s=-2$ and $s=0.5$.
We define the energy density as 
\begin{align}
    \varepsilon = \frac{E-E_\mathrm{min}}{E_\mathrm{max} -E_\mathrm{min} } \,,
\end{align}
with the energy $E$ and $E_\mathrm{min}$, $E_\mathrm{max}$ the smallest and largest many-body eigenenergies, respectively.

In the $s<0$ case, there is a dome-shaped structure typical for ergodic systems, while for $s=0.5$, this dome is accompanied by many low-entanglement states across the entire spectrum~\cite{Pancotti2020}.
A full classification of all the low-entanglement states remains open~\cite{Roy2020,Badbaria2024}.

\section{Quantum East Graph model and centrality measures}
\label{sec:qegraph}

\subsection{Quantum East Graph}
\label{sec:QEGraph}

Here, we describe how a graph can be associated to the QE Hamiltonian. We start by considering the model deep in the ergodic regime ($s<0$). 
In the limit $e^{-s}/\mu \to \infty$, the Hamiltonian~\eqref{eq:hamiltonian_qeast} can be rewritten as
\begin{align}
    \label{eq:hamiltonian_qeast_graph}
    H_\mathrm{QEG} = -\sum_\ell n_\ell \sigma^x_{\ell+1}\,,
\end{align}
where we have set the only remaining energy scale of the Hamiltonian to unity, i.e., $e^{-s} / 2 = 1$.

The Hamiltonian in \cref{eq:hamiltonian_qeast_graph} is purely off-diagonal, and can be directly identified with (minus) the adjacency matrix of a graph $g_\mathrm{QE}$ in which nodes $v_i \equiv \ket{i}$ correspond to configurations $\ket{i}$ in the Fock-space basis of particle-occupation eigenstates, and edges to transitions facilitated by the constraint.
Since the adjacency matrix $A$ of a graph is defined as~\cite{Newman2010}
\begin{align}
    A_{ij} = 
    \begin{cases}
1 & v_i\text{ and }v_j\text{ are connected}\\
0 & \text{otherwise}\,,
\end{cases}
\end{align}
we can identify 
\begin{align}
    (H_\mathrm{QEG})_{ij} = - A_{ij}\,.
\end{align}

Because of this identification, we call this limit of the QE model the quantum East graph model (QEG).
Note that the description of the QE model in terms of $g_\mathrm{QE}$ is basis-dependent.
The construction of the graphical representation of the graph $g_\mathrm{QE}$ is described in \cref{sec:appendix_construct_many_body_graphs}.

The Hamiltonian in \cref{eq:hamiltonian_qeast_graph} commutes with the operator $\sigma^x_L$, and thus the QE graph has a sub-graph symmetry corresponding to flipping the spin at the last lattice site (see also \cref{fig:qe_graph}).
The symmetry can be lifted by fixing appropriate boundary conditions, according to \cref{eq:HQE_actual}.
Furthermore, the QE graph is bipartite, which means that the nodes of the graph can be sorted into two groups whose nodes are only connected to the nodes of the other group.
As a consequence, all closed paths on the graph are of even length.

\subsection{Connection to graph Laplacians}

Another concept from graph theory that will be useful for our study is 
the \emph{Laplacian operator} $L$ of a graph $g$.
The graph Laplacian $L$ can be written as~\cite{Newman2010}
\begin{align}
\label{eq:laplacian_degree}
(L(g))_{ij} = \delta_{ij} D_i - (A(g))_{ij}
\,,\end{align}
where $A(g)$ is again the adjacency matrix of the underlying graph $g$ and $D_i = D(v_i)$ is the degree of connectivity, or degree
of node $v_i$, i.e., the number of edges of $g$ that involve $v_i$.

In the QE graph, the degree of a node $v_i$ is equal to the number of satisfied constraints in $\ket{i}$.
For the QE Hamiltonian in \cref{eq:hamiltonian_qeast_original}, 
\begin{align}
D_i =\sum\limits_{\ell=1}^{L-1} \mel{i}{n_\ell}{i}\,,
\end{align} and the Hamiltonian matrix elements can be written as
\begin{align}
\label{eq:hamiltonian_qeast_laplace}
        (H_\mathrm{QE})_{ij} = \frac{1}{2}(\delta_{ij} D_i - e^{-s} A_{ij})
\,.
\end{align}
Therefore, when $s=0$, the quantum East Hamiltonian~\eqref{eq:hamiltonian_qeast_original}
can be understood as the graph Laplacian of the QE graph.

\subsection{Graph centrality and hierarchy}
\label{sec:qegraph_centrality}

In the context of graph theory, graph centrality has long been studied in various different settings~\cite{Bavelas1950,Katz1953,Nieminen1974} to understand, e.g., how easily certain nodes can reach resources in a transport network~\cite{Guimera2005} or to identify important actors in a citation network~\cite{Liu2005}.
For a given graph $g$, measures of graph centrality can therefore highlight hierarchies, 
indicating which nodes can be considered important with respect to a particular, context-dependent property.
Measures of graph centrality assign 
a real value
to each node $v_i \in g$, where by convention larger values generally denote more central (important) nodes.
Over time, many different measures of graph centrality have been developed (see~\cite{Das2018} for a review) based on various properties of the underlying graph (e.g.,~spectrum, structure, distances).

In the context of the QE graph, we start with degree centrality ($\DC$)~\cite{Nieminen1974} as the simplest measure of centrality that captures some of the essential features of the QE graph.
$\DC$ is based on the degree of a node $D(v_i)$.
To compare $\DC$ with other measures of centrality it is often rescaled so that $\sum_i \DC(v_i) = 1$, hence 
\begin{align}
    \DC(v_i) = \frac{D_i}{\sum_j D_j}\,.
\end{align}
The basic premise of this measure is that nodes with higher degree will automatically also be more central, because
one can easily reach a larger number of other nodes (neighbors) 
if that node is directly connected to many others.

In the QE case with East boundary conditions, the degree of a node is $D_i=1+\sum_{\ell=1}^{L-1}\mel{i}{n_\ell}{i}$, i.e., it 
scales with the number of particles in the state, ${D_i \sim N_0 = \sum_\ell \mel{i}{n_\ell}{i}}$.
Therefore, the degree distribution for large system size $L$ is simply binomial $P_L(D_i = N_0 ) \sim \binom{L}{N_0}$, as already highlighted in~\cite{Roy2020}.
Hence, when $L$ is large, $D_i$ approximately follows a Gaussian distribution with mean over all nodes $\langle D_i \rangle = L/2$ and standard deviation $\sigma=\sqrt{L}/2$.

This fully characterizes the QE graph with respect to the $\DC$ centrality measure, which is independent of $s$.
However, while at $e^{-s}/\mu\to \infty$ the QE Hamiltonian is deep in the ergodic phase~\cite{Pancotti2020}, at finite $s$ the slow dynamics results from the competition of the allowed transitions and \emph{diagonal} terms.
Hence, the hierarchy as captured in the distribution of $\DC$ per se is not sufficient to explain the emergent slow dynamics for $s>0$.
We thus aim at identifying a centrality measure that captures also diagonal terms beyond only the connectivity of the Fock-space graph and can be related to the dynamical properties of the QE model at any finite $s$.

\subsection{Graph energy centrality}
\label{sec:graphenergy_centrality}

The Laplacian description of the QE model leads us to an alternative centrality measure called Laplacian centrality ($\LC$)~\cite{Qi2012}.
Laplacian centrality ($\LC$) is based on the more general concept of graph energy~\cite{Streitwieser1962,Junichi1976,Cvetkovic1980,Gutman1986,Gutman2001,Gutman2006}.

We define the Laplacian energy for a graph $g$ as the sum of squared eigenvalues of the Laplacian matrix~\cite{Qi2012} 
 $\GE_L(g) = \sum_n \lambda_n^2$.
The $\LC$ of a node $v_i$ is then defined as the relative change in Laplacian energy when the node is removed from the graph,
\begin{align}
\label{eq:laplacian_centrality}
    \LC_g(v_i) = \frac{\GE_L(g) - \GE_L(g \setminus v_i)}{\GE_L(g)}
\,,
\end{align}
where $g \setminus v_i$ denotes the graph
with the node $v_i$ and all edges connecting it to other nodes removed.
The $\LC$ also applies to  weighted graphs~\cite{Qi2012},
in which the adjacency matrix may contain 
arbitrary off-diagonal weights $A_{ij} = w_{ij}>0$ that indicate how strongly two nodes $v_i$ and $v_j$ are connected.%
\footnote{
The concept of graph energy centrality, defined slightly differently, has been used 
to study important players in social networks~\cite{Kamath2019}, epileptic brain regions~\cite{Nithin2021} and Covid-19 infection networks~\cite{Mahadevi2022}.
}

The expression for the graph energy $\GE_L$ can be rewritten
as~\cite{Qi2012}
\begin{align}
\label{eq:graph_energy_degree}
\GE_L(g) = \sum_n \lambda_n^2 = \sum\limits_{i} x_i^2 + \sum\limits_{i\neq j} w_{ij}^2
\,,
\end{align}
with 
$x_i = \sum_{v_j \in \mathcal{N}_g(v_i)} w_{ij}$ 
being the sum of weights leading to vertex $v_i$, where 
$\mathcal{N}_g(v_i)$ 
denotes the neighborhood of $v_i$, defined as all nodes which share edges with $v_i$ in $g$.
For unweighted graphs ($w_{ij}\in\{0,\,1\}$), we have $x_i = D_i$.
As a result, we may understand the $\LC$ as an extension of the earlier discussed degree centrality $\DC$ to weighted graphs.

Finally, we introduce the centrality measure 
that will be central to our analysis, applicable to Hamiltonians and inspired from the previous discussion of Laplacian centrality.
Given a Hamiltonian $H$, we define its graph energy as
\begin{align}
\label{eq:graph_energy_hamiltonian}
\GE(H) =\sum\limits_n E_n^2 = \Tr(H^2)
\,,\end{align}
with $E_n$ the eigenenergies of $H$. With $H$ expressed as a matrix $H_{ij}$
in a given basis, we obtain
\begin{align}
\label{eq:graph_energy_hamiltonian_matrix}
\GE(H)= \sum\limits_i H_{ii}^2 + \sum\limits_{i \neq j} H_{ij}^2 \,,
\end{align}
which makes apparent
the similarity to \cref{eq:graph_energy_degree}.

Using this definition of (Hamiltonian) graph energy, we introduce our main measure, the \emph{graph energy centrality} ($\GEC$) of a given state $\ket{i}$, defined as
\begin{align}
\label{eq:graph_energy_centrality}
    \GEC(\ket{i}) = \frac{\GE(H) - \GE(H \setminus \ket{i})}{\GE(H)}
\,,
\end{align}
where
\begin{align}
\label{eq:graph_energy_projected}
\GE(H \setminus \ket{i}) \equiv \GE[(\openone -\ket{i}\bra{i})H(\openone -\ket{i}\bra{i})] \,,
\end{align}
corresponds to the model after excluding configuration $\ket{i}$.

Since $\GE(H)$ depends on the full spectrum of the Hamiltonian, $\GEC$ is a centrality measure that indeed captures
the diagonal terms.
Further, we note that $\GE(H)$ and therefore also $\GEC$ is \textit{a priori} not invariant under spectral shifts $H \to H + c \openone$, with $c$ real.
Therefore, for a given Hamiltonian, before computing graph energies and the $\GEC$, we subtract the trace 
\begin{align}
    \label{eq:make_traceless}
    H \to H - \frac{\Tr(H)}{D}\,\openone\,,
\end{align}
where $D$ is the dimension of the subspace.
Through this step we ensure that $\GEC$ is invariant under unphysical arbitrary energy shifts.

We stress here that the new $\GEC$ measure is different from $\LC$ that was discussed above.
Whereas $\LC$ probes the structure of the graph $g$ through the adjacency matrix, our $\GEC$ captures the structure of the Hamiltonian matrix $H_{ij}$ in the Fock-space.
Even though $H_{ij}$ itself could be interpreted as a pure adjacency matrix of a more complex graph with self-loops (see e.g.,~\cite{Junichi1976,Acikmese2015,Vivek2023}) and appropriate weights, the $\GEC$ does not correspond to the $\LC$ of such a weighted graph, since the self-loop elements of a Laplacian for such a graph would simply cancel out.

For the purpose of our work, we say that the graph exhibits a \emph{hierarchy} with respect to the $\GEC$ when there exist many nodes with atypically large or small values of the measure, compared to the mean value.
Hence, we later will utilize the \emph{spread} of the distribution of graph energy centrality as a quantification of how hierarchical a given network is with respect to that measure.
This is different from the conventional notion of associating the value of a centrality measure itself with a node's importance.

In the next section, we show that $\GEC$ can be calculated efficiently for large systems and derive a closed form expression for the $\GEC$ distribution in the QE model.
In the following
\cref{sec:graph_detuning,sec:slow_dynamics_entanglement}, we demonstrate the usefulness of the graph energy centrality in predicting slow dynamics and non-ergodic spectral properties.

\subsection{Efficient calculation of graph energy centrality}
\label{sec:discussion}

We now introduce a method to efficiently compute $\GEC$ on large QE systems, beyond
what is accessible with exact diagonalization.
In contrast to, e.g.,~$\overline{c}(t)$, which is numerically expensive to calculate for larger system sizes, 
as it requires simulating the evolution of the quantum many-body state, 
the distribution of $\GEC$ can be efficiently computed for the models we consider.
Using $\GE(H) = \Tr(H^2)$, and \cref{eq:graph_energy_projected}, we can write 
\begin{align}
    \label{eq:ge_simplified}
    \GEC(\ket{i}) &= \frac{2 \mel{i}{H^2}{i} - \mel{i}{H}{i}^2}{\Tr(H^2)}\,.
\end{align}
For the class of models defined in \cref{eq:havenoname},
we write 
\begin{align}
    H_{\mathrm{QE},P}(s,z)&=-\frac{1}{2}e^{-s}A+\frac{1}{2}\Delta+z P\,,
\label{eq:genericH}
\end{align}
where $A$ is the adjacency matrix of the QE graph, and $\Delta$ the diagonal part of the same Hamiltonian. 
The equally diagonal detuning term $P$ depends on the protocol. In particular, for East boundary conditions, as presented in \cref{sec:introduction_model}, $A=\sigma^x_1+\sum_{\ell=1}^{L-1}n_{\ell} \sigma^x_{\ell+1}$ and $\Delta=\sum_{\ell=1}^{L-1}n_{\ell}-\gamma e^{-s}n_L+c\openone$, with $\gamma=\pm 1$ according to the sign choice in \cref{eq:HQE_actual} and using $c$ to make $H_{\mathrm{QE}, P}$ traceless according to \cref{eq:make_traceless}.
The terms in \cref{eq:ge_simplified} involve only diagonal contributions of $H$ or $H^2$, so that
we can express \cref{eq:ge_simplified} as 
\begin{align}
    \GEC(\ket{i}) 
    &=
    \frac{
        2 e^{-2s}\bra{i}A^2\ket{i} + \bra{i} \Delta  + 2z P \ket{i}^2
    }{
        \changes{4}\Tr(H_{\mathrm{QE}, P}^2)
    }\,,
\end{align}
with $\Tr(H_{\mathrm{QE}, P}^2) = e^{-2s} \Tr(A^2)+\Tr(\Delta +2zP)^2$.
We notice that $\mel{i}{A^2}{i} = \sum_{\ell=1}^{L-1}\mel{i}{n_{\ell}}{i}+1=N_0(i)-n_L(i)+1$.
Similarly, $\mel{i}{\Delta}{i}=N_0(i)-\gamma e^{-s}n_L(i)+c$, where we have defined
$N_0(i) = \sum_{\ell=1}^{L} \mel{i}{n_\ell}{i}$ and
$n_L(i)=\mel{i}{n_L}{i}$.
Notice that the degree distribution of the QE model which was discussed before in \cref{sec:qegraph_centrality} can be obtained in the same way.

As we show in \cref{sec:appendix_gec_pdf}, for most of the detuning protocols we study, 
 $\mel{i}{P}{i}$ is a function of a few occupation numbers of the state.
Therefore, in all these cases, $\GEC$ can take only a discrete set of polynomially many distinct values (in system size $L$), whose probabilities can be directly inferred from the probability distribution of configurations with certain occupations.
This means that the whole $\GEC$ distribution can be calculated exactly without the need to construct the exponentially large matrix of the Hamiltonian (see \cref{sec:appendix_gec_pdf} for the detailed derivation of the probability tables for each protocol).

The simplest example is the QE case ($z=0$). For a system of size $L$ we can define two variables $M=N_0(i)-n_L(i)$, taking values $M\in\{0,\,1,\ldots, L-1\}$, and $m=n_L(i)$ with values $m\in\{0,\,1\}$.
For each configuration, these variables determine the graph energy centrality, which can thus take only $2L$ different values
\begin{align}
&\GEC(M,m) = \nonumber\\
&
\frac{
    2 e^{-2s}(M+1)+(M+(1-\gamma e^{-s})m+c)^2
}{
    4\Tr(H_\mathrm{QE}^2)
}\,,
\label{eq:gec_qe}
\end{align}
and $\Tr(H_\mathrm{QE}^2) = 2^{L-3}[e^{-2s}(L+2)-\gamma e^{-s}(L+2c + 1)+\frac{1}{2}(L^2+(4c+1)L+4c^2)]$.
The probability that each of these values occurs is the probability that a given configuration has $m$ particles in the last site and $M$ in the remaining $L-1$ sites, namely
\begin{align}
    p(M,m)=p(M)p(m)=\frac{1}{2^{L}} \binom{L-1}{M}\,.
    \label{eq:gec_qe_prob}
\end{align}

\begin{figure}[t]
    \centering
    \includegraphics{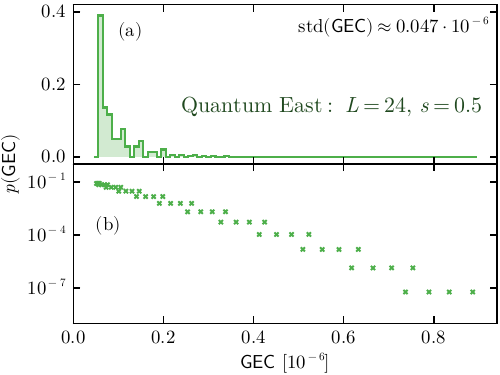}
    {\phantomsubcaption\label{fig:GEC_protocols_distributions_L24_histogram}}
    {\phantomsubcaption\label{fig:GEC_protocols_distributions_L24_logplot}}
    \caption{
        \subref{fig:GEC_protocols_distributions_L24_histogram}~Distribution of graph energy centrality $\GEC$ as a histogram for the QE model for a large system size (${L=24, s=0.5}$) in the $\gamma = +1$ symmetry sector.
        \subref{fig:GEC_protocols_distributions_L24_logplot}~Probability table of the same case (i.e., data before binning) as obtained from \cref{eq:gec_qe} and \cref{eq:gec_qe_prob}. Note the exponential decay in the larger-$\GEC$ tail.
    }
    \label{fig:GEC_protocols_distributions_L24}
\end{figure}

\Cref{fig:GEC_protocols_distributions_L24} shows the $\GEC$ of the QE model for a large system calculated using \cref{eq:gec_qe} and for $s=0.5$.
In \cref{fig:GEC_protocols_distributions_L24_histogram} we can see that the distribution is peaked around small values while \cref{fig:GEC_protocols_distributions_L24_logplot} highlights the discreteness of the distinct values.

Since we want to establish a connection between $\GEC$ distributions and other eigenstate measures and time-dependent quantities that typically require exact diagonalization (ED) of the Hamiltonian, for the remainder of this work we will consider small system sizes, amenable to ED, even though the $\GEC$ can be exactly computed for arbitrarily large systems.
An extended analysis of the properties of $\GEC$ distributions for large systems will be the subject of a follow-up work.

\section{Basis State Detuning Protocols}
\label{sec:graph_detuning}

\begin{table*}
    \begin{tabularx}{\linewidth}{|c|X|cr|l}
        \hline 
        & \centering{typical behavior} & detuning protocol $P$ & \\\hline\hline

        \noindent\parbox[c]{3cm}{nearest-neighbor interaction \\${\color{interactdetuning}\bullet}$}
        & \noindent\parbox[c]{4.5cm}{strongly enhanced slow dynamics, almost only sub-volume law eigenstates}
        & \noindent\parbox[c]{6cm}{ $$P_\mathrm{interact} = \sum\limits_{\ell = 1}^{L-1} n_\ell n_{\ell+1}$$}
        & \includegraphics[width=0.2\linewidth,margin=0pt 1ex 0pt 1ex,valign=m]{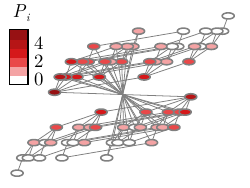}
        \\\hline

        \noindent\parbox[c]{3cm}{local operator\\${\color{localdetuning}\bullet}$}

        & \noindent\parbox[c]{4.5cm}{slightly enhanced slow dynamics, mix of sub-volume and volume law ekgenstates/less volume law state than $P=0$}
        & \noindent\parbox[c]{6cm}{$$P_\mathrm{local} = n_{L / 2}$$}
        & \includegraphics[width=0.2\linewidth,margin=0pt 1ex 0pt 1ex,valign=m]{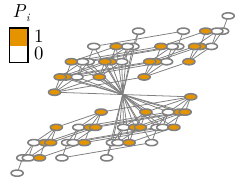}
        \\\hline

        \noindent\parbox[c]{3cm}{QE detuning\\${\color{nodetuning}\bullet}$}
        & \noindent\parbox[c]{4.5cm}{slow dynamics ($s=0.5$), mix of sub-volume and volume law eigenstates} 
        & \noindent\parbox[c]{6cm}{$$P = 0 \quad \left(\Delta = \sum_{\ell=1}^{L-1} n_\ell \right)$$}
        & \includegraphics[width=0.2\linewidth,margin=0pt 1ex 0pt 1ex,valign=m]{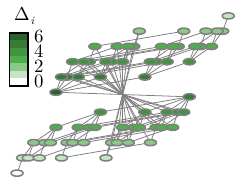}
        \\\hline

        \noindent\parbox[c]{3cm}{density detuning \\${\color{fillingdetuning}\bullet}$}
        & \noindent\parbox[c]{4.5cm}{reduced slow dynamics, many volume law eigenstates}
        & \noindent\parbox[c]{6cm}{ $$P_\mathrm{filling} = \sum_{i; N_0(i)=L/2} \ket{i}\bra{i}$$ } 
        &\includegraphics[width=0.2\linewidth,margin=0pt 1ex 0pt 1ex,valign=m]{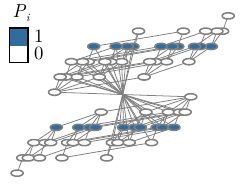}
        \\\hline

        \noindent\parbox[c]{3cm}{greedy optimization \\${\color{greedydetuning}\bullet}$}
        & \noindent\parbox[c]{4.5cm}{almost no slow dynamics, volume law eigenstates}
        & \noindent\parbox[c]{6cm}{ $$P_\mathrm{greedy} = \sum_k c^{(k)}\ket{\alpha_k}\bra{\alpha_k}$$ }
        &\includegraphics[width=0.2\linewidth,margin=0pt 1ex 0pt 1ex,valign=m]{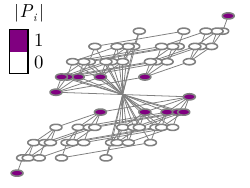}
        \\\hline
    \end{tabularx}
    \caption{
        Summary of the different detuning protocols.
        We show both the general form of the operator $P$ implementing the detuning protocol as an operator and highlight its graphical structure by showing its expectation values $P_i = \mel{i}{P}{i}$ in the states $\ket{i}$ corresponding to the QE graph nodes, for $L=6$.
        For the undetuned system $P=0$, we show the diagonal detunings in the QE Hamiltonian $\Delta_i = \mel{i}{\Delta}{i}$ disregarding boundary terms (corresponding to $\gamma = +1, s=0$).
        Colors in the leftmost column indicate how results for each protocol will be represented throughout this work.
    }
    \label{tab:detuning_protocols}
\end{table*}

Our working hypothesis is that slow dynamics emerges due to hierarchies between nodes on the graph as measured by the spread of $\GEC$~Eq.~\eqref{eq:graph_energy_centrality}.
Because this measure is sensitive to diagonal terms in the Hamiltonian, it can be affected by a relative \emph{detuning} of the states.

In this section, we thus introduce four detuning protocols, 
with respect to a specific
QE Hamiltonian at a fixed value of $s=0.5$, 
and characterize the resulting five Hamiltonians, including the undetuned QE model, via their $\GEC$.
Using density autocorrelations, we will demonstrate that by employing these protocols, we can tune the dynamical behavior of the system.

\subsection{Detuning protocols}
\label{sec:graph_detuning_protocols}

\begin{figure}[t]
    \centering
    \includegraphics{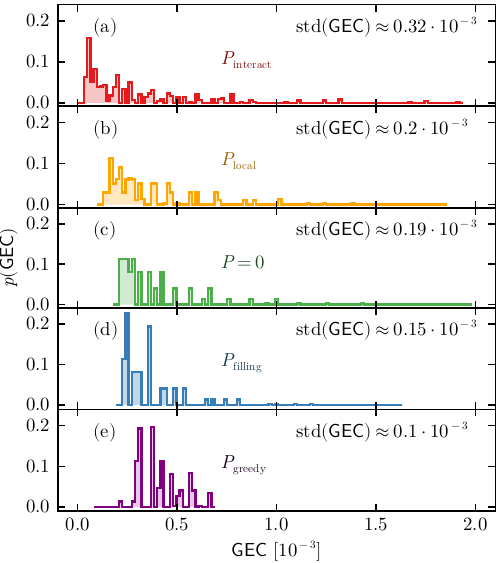}
    {\phantomsubcaption\label{fig:GEC_protocols_distribution_interact}}
    {\phantomsubcaption\label{fig:GEC_protocols_distribution_local}}
    {\phantomsubcaption\label{fig:GEC_protocols_distribution_nothing}}
    {\phantomsubcaption\label{fig:GEC_protocols_distribution_filling}}
    {\phantomsubcaption\label{fig:GEC_protocols_distribution_greedy}}
    \caption{
        Distribution of graph energy centrality $\GEC$ for all different detuning protocols \subref{fig:GEC_protocols_distribution_interact}-\subref{fig:GEC_protocols_distribution_greedy} as defined in \cref{tab:detuning_protocols} ($L=12, s=0.5$).
        \subref{fig:GEC_protocols_distribution_nothing}~$P=0$ corresponds to the standard QE model.
    }
    \label{fig:GEC_protocols_distributions}
\end{figure}

\begin{figure*}[t]
    \centering
    \includegraphics{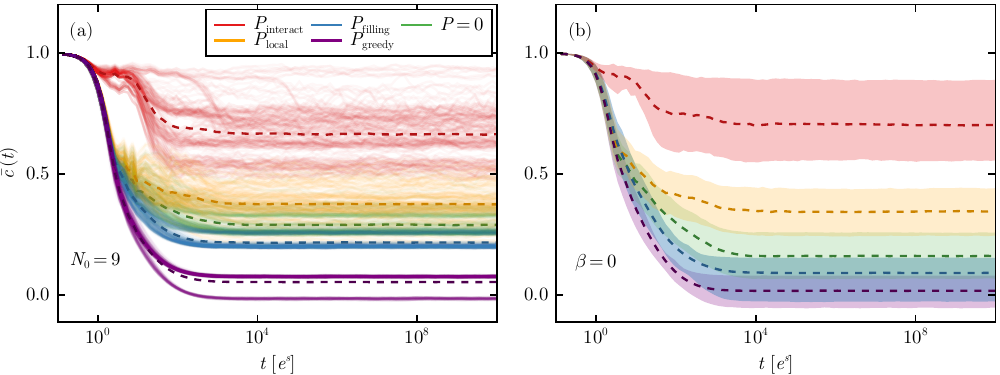}
    {\phantomsubcaption\label{fig:autocors_N9}}
    {\phantomsubcaption\label{fig:autocors_mean}}

    \caption{
        Normalized density autocorrelation functions for the different detuning protocols (see \cref{tab:detuning_protocols} for a legend).
        In \subref{fig:autocors_N9} we show the time-averaged autocorrelation functions $\overline{c}(t)$ in the detuned quantum East models for $L=12$, $s=0.5$ for initial states with an initial particle number $N_0=9$.
        We also show the average over the ensemble of all selected initial states as a dashed line in the corresponding color.
        In \subref{fig:autocors_mean} we show the mean of $\overline{c}(t)$ over all initial states (i.e., the infinite temperature average $\beta = 0$).
        The shaded area around each dashed line shows the central \qty{68}{\percent} region of the distribution of $\overline{c}(t)$ across all initial state.
    }
    \label{fig:aucorrs_protocols}
\end{figure*}

We define the detuned quantum East Hamiltonian as
\begin{align}
\label{eq:detuned_qeast_hamiltonian}
H_{\mathrm{QE},P} = H_\mathrm{QE} + z P
,\end{align}
where $P$ is an operator that acts on all Fock states that we want to affect with the detuning, and $z$ controls the detuning strength.
We fix the QE Hamiltonian at ${s=0.5}$, as this corresponds to the most interesting regime, where slow dynamics are already present.
In all numerical simulations, we use a detuning strength of $z = 1$.

In the following, we describe in detail the various detuning protocols considered in this work, which exhibit broadly different dynamical and eigenstate properties, summarized in~\cref{tab:detuning_protocols},
and differ also in their $\GEC$ distributions,
shown in \cref{fig:GEC_protocols_distributions}.
Notice that in~\cref{fig:GEC_protocols_distributions} and~\cref{tab:detuning_protocols}, the protocols are ordered by decreasing spread of the $\GEC$ distribution.

Our first way of detuning the QE system uses nearest-neighbor density-density interactions
\begin{align}
\label{eq:detuning_projector_interactions}
    P_\mathrm{interact} = \sum_{\ell = 1}^{L-1} n_\ell n_{\ell+1}
\,.\end{align}
In \cref{tab:detuning_protocols} we show the structure of $P_\mathrm{interact}$ on the QE graph.
Every allowed transition will necessarily be detuned due to the neareast-neighbor interactions ($\bullet \circ \leftrightharpoons \bullet \bullet$). Therefore, we expect slower dynamics than in the reference QE model.
Results for the interaction detuning protocol are shown in {\color{interactdetuning}red~${\bullet}$} throughout this manuscript.

Local operators can also have a strong impact on the constrained dynamics of the finite QE model.
The second detuning protocol applies a single local operator $n_{L / 2}$ in the middle of the chain
\begin{align}
\label{eq:detuning_single_field}
P_\mathrm{local} = n_{L / 2} 
.\end{align}
As the operator $n_{L / 2}$ detunes exactly half of the states on the Fock-space graph, it is expected to have a strong impact on the slow dynamics as well.
We indicate results for the local detuning protocol by {\color{localdetuning}orange~${\bullet}$} color.

Instead of using many-body projection operators to detune the dynamics, we can also devise detuning protocols where we target individual states on the Fock graph.
One possible way is to detune all states with a fixed given particle number $N_0$.
We focus on the case $N_0=L/2$.
On the QE graph, such a protocol 
 has the effect of a barrier splitting the graph into two regions (see \cref{tab:detuning_protocols}).
We write this detuning protocol in the following way
\begin{align}
\label{eq:detuning_density}
P_\mathrm{filling} = \sum_{i; N_0(i)=L/2} \ket{i}\bra{i}
.\end{align}
In contrast to the previously introduced detuning protocols, this protocol 
reduces the degree of non-ergodicity as compared to the QE model.
In the following, we denote the density detuning protocol in {\color{fillingdetuning}blue~${\bullet}$}.

Lastly, we introduce a detuning protocol based on a \enquote{greedy} algorithm that uses graph energy centrality as a heuristic measure to detune individual nodes on the Fock-space graph 
with the goal of minimizing the the standard deviation of the $\GEC$ distribution.

Our algorithm proceeds by repeatedly choosing the state $\ket{i}$ that corresponds to the largest outlier in $p(\GEC)$ (which in the cases studied here is always the largest $\GEC$ value) and detunes it such that the overall spread $\mathrm{std}(\GEC)$ is reduced.

More precisely, starting
from $\tilde{H}^{(0)} = H_\mathrm{QE}$ 
we iteratively define a sequence of Hamiltonians
\begin{align}
    \label{eq:greedy_algorithm}
    \tilde{H}^{(k+1)} &= \tilde{H}^{(k)} + z c^{(k)}\ket{\alpha_k}\bra{\alpha_k}\,,\quad \text{with}\nonumber\\[5pt]
    \ket{\alpha_k} &= \argmax_{\ket{i}} \,\,\GEC_{\tilde{H}^{(k)}}(\ket{i})\,.
\end{align}
We obtain $c^{(k)}$ from
\begin{align}
    c^{(k)} = 
    \sgn[ 
        \mathrm{std}(\GEC_{\tilde{H}^{(k+1), -}})
        -
        \mathrm{std}(\GEC_{\tilde{H}^{(k+1), +}})
    ]\,,
\end{align}
which corresponds to selecting, from two trial Hamiltonians $\tilde{H}^{(k+1),c} = \tilde{H}^{(k)} + cz \ket{\alpha_k}\bra{\alpha_k}$ with $ c \in \{-1, +1\}$, the one with the smaller $\GEC$ spread.
This procedure is repeated up to $k_{\max}= \binom{L}{L / 2}$ times, such that the total number of detuning steps is the same as in the $P_\mathrm{filling}$ protocol, i.e., $k_\mathrm{max} = \Tr(P_\mathrm{filling})$.

When applied to the QE system, the algorithm in practice converges to a period-two cycle, repeatedly detuning the same node by $+1$/$-1$, before reaching $k_\mathrm{max}$.
A further improvement could be achieved by adjusting $z$ 
during the process. Exploring this is left for future work.

Eventually, we may write the full detuning protocol as
\begin{align}
\label{eq:detuning_greedy_alg}
P_\mathrm{greedy} = \sum_k c^{(k)}\ket{\alpha_k}\bra{\alpha_k}
\,.\end{align}
The sum can contain repeated terms since we allow states to be detuned multiple times.
Notice that we also allow for a negative detuning of states.
Data from the detuning protocol derived from our greedy optimization algorithm is shown in {\color{greedydetuning}purple~$\bullet$}.

Figure~\ref{fig:GEC_protocols_distributions} compares the $\GEC$ distributions for the five protocols (including the original QE model, which corresponds to $P=0$) for a system of size $L=12$ (see \cref{sec:appendix_gec_pdf} for details about computing the distributions for the detuned protocols).
In \cref{fig:GEC_protocols_distribution_interact,fig:GEC_protocols_distribution_local}, we observe how the distributions for the $P_\mathrm{interact}$ and $P_\mathrm{local}$ protocols accumulate more weight in the tails at large $\GEC$ values and a larger spread as quantified by their standard deviation, while in \cref{fig:GEC_protocols_distribution_filling,fig:GEC_protocols_distribution_greedy} the distributions of the $P_\mathrm{filling}$ and $P_\mathrm{greedy}$ protocols have smaller tails and a lower standard deviation, compared to the case of the original QE model shown in \cref{fig:GEC_protocols_distribution_nothing}. 
We will see next that the behavior of time-dependent density autocorrelations is consistent with this behavior in the $\GEC$ distributions.

Before moving on, note that the protocols clearly differ in how many nodes are detuned. In all cases, we are affecting exponentially or at least binomially
many states. How these different ratios of detuned versus non-detuned nodes affect the eventual ergodicity as system size increases is beyond the scope of this study.

\subsection{Effect of diagonal detuning on density autocorrelations}
\label{sec:graph_detuning_diagonal}

\begin{figure}[t]
    \centering
    \includegraphics{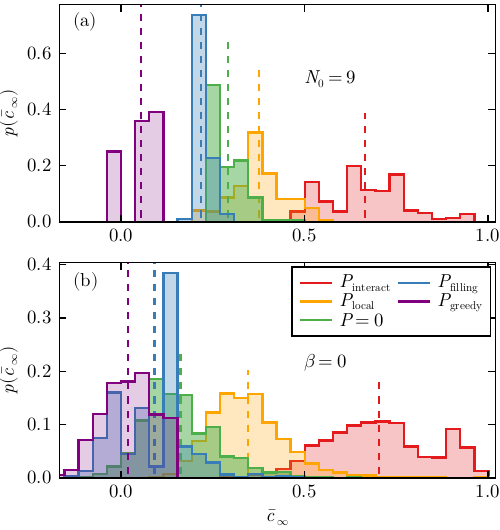}
    {\phantomsubcaption\label{fig:aucorrs_protocols_hist_N=9}}
    {\phantomsubcaption\label{fig:aucorrs_protocols_hist_all}}

    \caption{
        Distributions of the long-time value of the time-averaged autocorrelation function $\overline{c}_\infty$ for \subref{fig:aucorrs_protocols_hist_N=9} initial states with $N_0 = 9$ and \subref{fig:aucorrs_protocols_hist_all} for all initial states in the detuned QE models, i.e., the infinite-temperature average ($L=12$, $s=0.5$).
        Different colors denote the different detuning protocols (see \cref{tab:detuning_protocols} for a summary).
        We also indicate the mean value of these distributions with a vertical dashed line in the respective color.
        For the distribution in the case of $P_{\rm greedy}$, see the discussion in Sec.~\ref{sec:introduction_model}.
    }
    \label{fig:aucorrs_protocols_hist}
\end{figure}

\begin{figure}[t]
    \includegraphics[width=\linewidth]{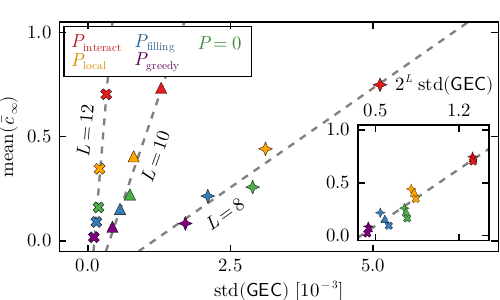}

    \caption{
        Long-time value of the time-averaged density autocorrelation function $\overline{c}_\infty$ averaged over all initial states in the computational basis versus the spread of graph energy centrality $\GEC$ as measured by the standard deviation of the $\GEC$ distribution for the different detuning protocols ($s=0.5$).
        Colors indicate the different detuning protocols, as given in the legend, and different symbols indicate different systems sizes of $L=8,10,12$ (stars, triangles, crosses).
        The inset shows the same data but we rescale $\mathrm{std}(\GEC)$ with Hilbert space dimension $2^L$ in order to compare the behavior of different system sizes.
        Dashed lines are linear fits to the respective datasets and serve as a guide to the eye.
    }
    \label{fig:GECs_protocols_std_vs_mean}
\end{figure}

So far, we have introduced the detuning protocols and discussed the associated distributions of the graph energy centrality, $p(\GEC)$. We now turn to the behavior of density autocorrelations.

Slow dynamics manifests itself in time-averaged autocorrelation functions $\overline{c}(t)$ that do not decay, or do so slower than usual~\cite{Lan2018}, see \cref{sec:introduction_model}.
For instance, in the regime around $s=0.5$ of the quantum East model, we observe slow dynamics in $\overline{c}(t)$ [see \cref{fig:intro_figures_autocors}].

In \cref{fig:autocors_N9}, we show $\overline{c}(t)$ for the different detuning protocols for the ensemble of initial states with an initial occupation $N_0 = 9$  with $L=12$ and $s=0.5$.
As expected, the detuning protocols $P_\mathrm{interact}$ and $P_\mathrm{local}$ enhance the already slow dynamics compared with the original undetuned QE model.
In contrast, we observe that the $P_\mathrm{filling}$ and $P_\mathrm{greedy}$ protocols result in a suppression of the slow dynamics.
\changes{The two distinct late-time values of $\overline{c}_\infty$ in \cref{fig:autocors_N9} for the $P_\mathrm{greedy}$, $P_\mathrm{filling}$ and $P=0$ protocols are connected to the initial occupation of the last lattice site (see also Ref.~\cite{Pancotti2020}).}
The same trend is evident in the infinite-temperature average over all $N_0$
shown in \cref{fig:autocors_mean}.
To corroborate the data from \cref{fig:aucorrs_protocols}, in \cref{fig:aucorrs_protocols_hist}, we show the distribution of the long-time values  $\overline{c}_\infty = \overline{c}(t = 10^{10} e^{-s})$ 
with fixed $N_0=9$ in \cref{fig:aucorrs_protocols_hist_N=9} and for all states in the computational basis in \cref{fig:aucorrs_protocols_hist_all}, and compare their behavior for the different detuning protocols.

To explain the change in the dynamical properties of the system we may work backwards through the detuning protocols: 
By construction, the $P_\mathrm{greedy}$ protocol aims to reduce hierarchies on the graph as measured by the spread of the graph energy centrality.
As we observe, this protocol also succeeds in suppressing the slow dynamics completely, indicating that our notion of graph theoretical hierarchy and the slow dynamics are indeed linked.

From the greedy algorithm we conclude that reducing hierarchies between states can suppress the slow dynamics.
In the $P_\mathrm{filling}$ detuning protocol, we introduce an energy barrier between nodes with filling $L/ 2 - 1$ and $L / 2$, while decreasing the energy difference between the nodes with filling $L / 2$ and $L / 2 + 1$, see the sketch of the corresponding graph in \cref{tab:detuning_protocols}.
We hypothesize that the reason for the reduction of the slow dynamics in this detuning protocol can be explained effectively by a decoupling of the nodes with $D_i < L / 2$ and $D_i \ge L / 2$.
These two subgroups only weakly couple to each other due to the energy barrier%
\changes{,
which results in eigenstates being mainly supported on only one of the subgroups. As discussed in the next section, our results indicate that each group of eigenstates becomes more extended than the eigenstates of the original $P=0$ protocol.
For more details on the $P_\mathrm{filling}$ protocol, see \cref{sec:appendix_filling}.
}

In contrast to the first two protocols, $P_\mathrm{local}$ does not reduce hierarchies between the basis states with respect to $\GEC$, but enhances them, see \cref{fig:GEC_protocols_distribution_local}.
From \cref{tab:detuning_protocols}, we see that $P_\mathrm{local}$ detunes exactly half of all product states in the graph.
Due to the local structure of the detuning, combined with the local structure of the graph's facilitated transitions, $P_\mathrm{local}$ detunes states on the graph such that 
several disjoint groups of undetuned nodes emerge.%
\footnote{
    An inspection of the graph's structure for $L \leq 22$ shows that there are $\lfloor (L+1) / 2 \rfloor+1$ such groups.
}
Transitions between these groups are significantly suppressed due to the detuning, enhancing the slow dynamics.

Lastly, we discuss the $P_\mathrm{interact}$ protocol which has an even larger spread of $\GEC$ than the local detuning protocol.
States targeted in this detuning protocol are generally those nodes which already have a higher filling and therefore also a higher degree of connectivity, see the discussion in \cref{sec:qegraph_centrality} or \cref{tab:detuning_protocols}.
In this way, $P_\mathrm{interact}$ enhances hierarchies on the graph strongly.

In the next step, we further elucidate the connection of $\GEC$ to the slowness of the overall dynamics by discussing the degree of hierarchy encoded in the $p(\GEC)$ distributions.

In order to establish a more quantitative connection between the $\GEC$ distribution and the ergodicity of the dynamics, in \cref{fig:GECs_protocols_std_vs_mean} we plot the average $\overline{c}_\infty$ value vs. the standard deviation $\mathrm{std}(\GEC)$ for the different protocols.
The figure indicates a positively correlated trend for all system sizes, indicating that the spread of the $\GEC$ distribution and the ensemble averaged $\overline{c}_\infty$ are indeed related.
We also observe that, by rescaling $\mathrm{std}(\GEC)$ with the Hilbert space dimension, different system sizes show highly similar behavior, which can be seen in the inset of \cref{fig:GECs_protocols_std_vs_mean}.

In this section we have presented evidence for the two main results of our work. First, detuning states in a systematic way and guided by the here defined $\GEC$ provides control over the degree of non-ergodicity.
Second, a systematic characterization arises involving graph energy centrality and its distribution.
The stronger the hierarchy is among the basis states with respect to this measure, the more significant the degree of non-ergodicity is.

\section{Eigenstate structure}
\label{sec:slow_dynamics_entanglement}

We now turn to eigenstate properties in the detuned models, covering spectral statistics, as probed by the gap ratio, and eigenstate entanglement entropy.

\subsection{Spectral statistics}

\begin{figure}[t]
    \centering
    \includegraphics[width=\linewidth]{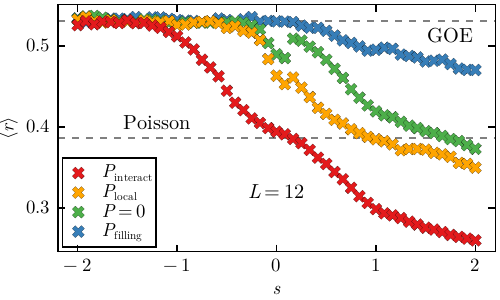}
    \caption{
    Spectral statistics for different detuning protocols as a function of $s$ for $L=12$.
    We show the gap ratio $\langle r \rangle$ averaged over 50 realizations with varying chemical potential $\mu = 1 + \delta \mu$ for $\delta \mu \in [-0.05, 0.05]$ for the $P_\mathrm{interact}$, $P_\mathrm{local}$ and $P_\mathrm{filling}$ detuning protocols and compare with the undetuned QE system ($P = 0$) [see also \cref{fig:intro_figures_gapratio}].
    }
    \label{fig:gapratio_protocols}
\end{figure}

\begin{figure}[t]
    \centering
    \includegraphics[width=\linewidth]{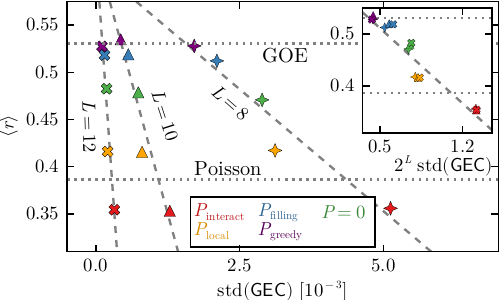}
    \caption{
    Spectral statistics versus $\mathrm{std}(\GEC)$.
    We show data for the gap ratio $\langle r \rangle$ averaged over 50 samples with varying chemical potential $\mu = 1 + \delta \mu$ for $\delta \mu \in [-0.05, 0.05]$ at fixed $s=0.5$.
    Colors indicate the different detuning protocols, as given in the legend, and different symbols indicate different systems sizes of $L=8,10,12$ (stars, triangles, crosses).
    The inset shows the same data but we rescale $\mathrm{std}(\GEC)$ with Hilbert space dimension $2^L$ instead of rescaling with $10^{-3}$.
    Dashed lines are linear fits to the respective datasets and serve as a guide to the eye and dotted lines indicate the respective GOE and Poisson values for the gap ratio.
    }
    \label{fig:gapratio_vs_std_gec}
\end{figure}

In \cref{fig:gapratio_protocols}, we show the dependence of the gap ratio on the parameter $s$ for the $P_\mathrm{filling}$, $P_\mathrm{local}$ and $P_\mathrm{interact}$ detuning protocols.
We do not show the results for the $P_\mathrm{greedy}$ protocol in \cref{fig:gapratio_protocols} because 
$P_\mathrm{greedy}$ is designed to  yield values
of $\langle r\rangle$ consistent with GOE (convergence may depend on $s$, though).

The gap-ratio average is obtained  from an ensemble of 50 realizations, where we randomly sample the chemical potential strength from a uniform distribution of $\mu = 1 + \delta \mu$ with $\delta \mu \in [-0.05, 0.05]$.
We observe that at large negative $s$ all cases behave according to random matrix theory, whereas as $s$ is increased, they all exhibit a deviation from random matrix statistics beyond a value of $s$ that depends on the detuning protocol.
These results are consistent with our observations from the density autocorrelations, as illustrated in \cref{fig:aucorrs_protocols}: 
the extent of the region with GOE statistics grows as we go from $P_\mathrm{interact}$ to $P_\mathrm{local}$, to the standard QE model and then to $P_\mathrm{filling}$.
Note that for $P_\mathrm{local}$, as for the undetuned case $P=0$, $\langle r \rangle$ shows a small minimum around $s=0$ because the symmetry on the last lattice site discussed in \cref{sec:introduction_model} is not lifted, different to what happens for the other detuning protocols.
\changes{
Note that at large $s$, we observe that $\langle r \rangle$ drops below the Poisson value for the gap ratio.
This can be attributed to the opening of bands in the spectrum that was discussed in \cref{sec:introduction_model}, which leads to a system that is nearly particle conserving.
This introduces unresolved near-symmetries into the system whose artifacts are picked up by the gap ratio.
}

In \cref{fig:gapratio_vs_std_gec} we show that the gap ratio and $\mathrm{std}(\GEC)$ are correlated across all the detuned models, indicating that the degree of non-ergodicity captured by the spectral statistics is also reflected in the hierarchy between nodes on the graph.

\subsection{Entanglement Entropy}
\label{sec:eigenstate_entanglement}

\begin{figure}[t]
    \includegraphics[width=\linewidth]{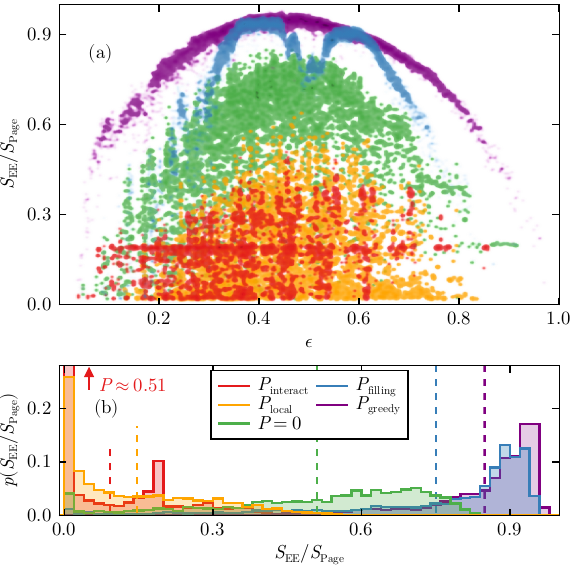}
    {\phantomsubcaption\label{fig:EEs_protocols_spectrum}}
    {\phantomsubcaption\label{fig:EEs_protocols_hist}}

    \caption{
      Spectrum of the half-chain entanglement entropy $S_\mathrm{EE}$ for the QE Hamiltonians ($L=12$, $s=0.5$) with different detuning protocols, indicated by different colors [see the legend in \subref{fig:EEs_protocols_hist}].
      The entanglement entropy is rescaled by the Page value $S_\mathrm{Page}$.
      \subref{fig:EEs_protocols_spectrum}~Entanglement entropy $S_\mathrm{E E} / S_\mathrm{Page}$ as a function of energy density $\epsilon$.
      \subref{fig:EEs_protocols_hist}~Distribution of entanglement entropy $p(S_\mathrm{E E} / S_\mathrm{Page})$.
      The mean of the distributions are indicated by vertical dashed lines in the corresponding colors.
      For the protocol $P_\mathrm{interact}$, there is a large amount of states with low entanglement $S_\mathrm{EE}/S_\mathrm{Page} < \qty{2}{\percent}$, indicated by the value in~\subref{fig:EEs_protocols_hist}.
      }
    \label{fig:EEs_protocols}
\end{figure}

\begin{figure}[t]
    \centering
    \includegraphics[width=\linewidth]{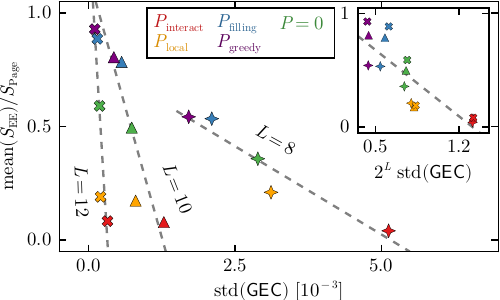}
    \caption{
    Mean entanglement entropy $S_\mathrm{EE}$ of the middle \qty{50}{\percent} eigenstates versus $\mathrm{std}(\GEC)$ at $s=0.5$.
    Colors indicate the different detuning protocols, as given in the legend, and different symbols indicate different systems sizes of $L=8,10,12$ (stars, triangles, crosses).
    The inset shows the same data but we rescale $\mathrm{std}(\GEC)$ with Hilbert space dimension $2^L$ instead of rescaling with $10^{-3}$.
    Dashed lines are linear fits to the respective datasets and serve as a guide to the eye.
    }
    \label{fig:EE_vs_std_gec}
\end{figure}

The changes in the dynamical properties of the system as the detuning is varied should also be reflected in the distribution of entanglement entropy in the eigenstates. 
We focus on $s=0.5$.
In \cref{fig:EEs_protocols}, we analyze 
the half-chain entanglement entropy from \cref{eq:entanglement_entropy} for the eigenstates of each system.

\Cref{fig:EEs_protocols_spectrum} shows that the various detuning protocols exhibit vastly different shapes of the eigenstate entanglement as a function of the energy density.
The detunings $P_\mathrm{greedy}$ and $P_\mathrm{filling}$ indeed show dome-like features in the energy-dependence of the entanglement entropy, which hints at the existence of ergodic eigenstates. 
These domes are narrow and follow volume-law scaling (see, e.g.,~\cite{DAlessio2016}). 
In contrast, the eigenstates in the regular QE model ($P=0$) and the detuning protocols $P_\mathrm{local}$ and $P_\mathrm{interact}$ host many eigenstates with atypically low entanglement.

\Cref{fig:EEs_protocols_spectrum} represents a projection on the entropy-energy plane of the eigenstates (probability) distribution. 
By additionally examining the marginal distribution $p(S_\mathrm{EE}/S_\mathrm{Page})$ across eigenstates in \cref{fig:EEs_protocols_hist}, we observe that the distributions of $S_\mathrm{EE}$ also reflect our previous analysis.
Protocols that show a dome-like feature, such as $P_\mathrm{greedy}$ and $P_\mathrm{filling}$, show mostly high-entanglement eigenstates, while the absence of the dome feature is accompanied with a fraction of states with atypically low entanglement.
Particularly, while $P=0$ still hosts states at typically large values of entanglement [see also \cref{fig:intro_figures_EEs}], nearly all states for $P_\mathrm{local}$ have atypically low $S_\mathrm{EE}$.
This effect is even stronger for the $P_\mathrm{interact}$ protocol.
In this case, and for $L=12$, about \qty{51}{\percent} of all eigenstates have an entanglement entropy that is smaller than \qty{2}{\percent} of $S_\mathrm{Page}$.

\Cref{fig:EEs_protocols_spectrum} reveals also finer details about the respective eigenstate entanglement structures.
In particular, for the $P_\mathrm{filling}$ protocol, which shows dome-like features, there is additionally a pronounced dip in the typical value of $S_\mathrm{EE}$ around $\epsilon = 0.5$.
We attribute this to the local structure of $P_\mathrm{filling}$, which detunes basis states that separate the low-filling  from the high-filling basis states [see \cref{tab:detuning_protocols} 
and Fig.~\ref{fig:GEC_protocols_distribution_filling}].
Through $P_\mathrm{filling}$ we have introduced an energy barrier between these two partitions of the graph, 
which produces a distribution with two maxima of $S_\mathrm{EE}$.
The reason is the following:
generally, there is a correlation between particle number and energy density of eigenstates in these models.
 Eigenstates for $P_\mathrm{filling}$ have large support either on basis states with $N_0<L/2$ 
 or $N_0>L/2$.
Hence, we have essentially decoupled both regimes. 

For $P_\mathrm{local}$, the extensive number of detuned groups of states, that have already been discussed in \cref{sec:graph_detuning_diagonal}, can be used to understand the existence of many eigenstates with atypically low entanglement. 

For the interaction protocol, $S_\mathrm{EE}$ exhibits clear band-like structures.
Their energy separation is given by $\Delta \varepsilon \approx 1/L$.
The influence of the nearest-neighbor interaction has an effect similar to the one of the chemical potential term that was discussed in \cref{sec:introduction_model}, leading to the formation of bands in the finite system.
Moreover, there are many eigenstates at $S_\mathrm{EE} \approx \log(2)$ [horizontal structure in the $S_\mathrm{EE}$ vs $\varepsilon$ plot, \cref{fig:EEs_protocols_spectrum}], indicating the presence of many two-particle
resonances, similar to many-body localized phases 
(see, e.g.,~\cite{Lim2016}).

In analogy to the other ergodicity indicators, we investigate the correlation of the average $S_\mathrm{EE}$ for the central \qty{50}{\percent} eigenstates with the standard deviation of the $\GEC$.
The results, presented in \cref{fig:EE_vs_std_gec}, demonstrate that a correlation between our centrality measure and the $S_\mathrm{EE}$ exist in this case as well.
Overall, the observations drawn from the eigenstate entanglement are consistent with those made in the analysis of autocorrelations.

\section{Summary and Outlook}
\label{sec:summary}

In this work we developed the hypothesis that hierarchy among the states in a given computational basis can be a predictor of slow dynamics in a quantum system.
Focusing on the quantum East (QE) model, we introduced
the \emph{graph energy centrality} ($\GEC$) measure 
to define and investigate such hierarchies in a family of QE models.
Our construction of $\GEC$ is inspired by the resemblance between the QE Hamiltonian and the Laplacian operator of its connectivity graph in the computational basis.

To quantify hierarchy in the system we investigated the spread of the $\GEC$ distribution.
Next, through the properties of the $\GEC$, we recognized that hierarchy in the system critically depends on the relative detuning of states and therefore, we investigated different detuning protocols to support our original hypothesis.

We introduced four ways of detuning states, and compared them to a reference QE model in its slow-dynamics regime.
First, we showed that through a greedy, iterative detuning protocol that aims at incrementally reducing the spread of $\GEC$,
we can completely restore ergodic behavior, providing strong evidence that $\GEC$-based hierarchy and ergodicity are linked.
We also considered another three examples of detuning protocols, for which we demonstrated that slow dynamics is either enhanced or suppressed,
consistent
with their respective $\GEC$ distributions.
Furthermore, we showed and discussed how the eigenstate structure of the different detuned systems is linked to their spread of $\GEC$ values.
The results presented in this work are obtained for finite systems and we do not attempt to draw conclusions on the persistence of slow dynamics in the thermodynamic limit.
This is left for future work.

On a technical level, we demonstrated that $\GEC$ can be efficiently computed for system sizes much beyond the reach of exact diagonalization.
While there is further need to improve the physical interpretation of $\GEC$ distributions, it is noteworthy that essentially
all other measures for quantum chaos, such as the inverse participation ratio, entanglement entropy, and gap ratio,
require the full diagonalization of the Hamiltonian and are thus only computable for small systems.

Our work opens an array of future research directions.
We have provided an interpretation for the emergence of slow dynamics in the context of the QE model and the natural question is to ask how widely applicable the concept of $\GEC$-based hierarchy is to other models.
A natural starting point are those models that are closely related to Laplacians such as, e.g., the triangular lattice gas model~\cite{Lan2018,Royen2024}.
Ultimately,
one would want to connect 
as well to models that can be realized in concrete experiments 
(see, e.g.,~\cite{ValenciaTortora2024} for proposals on how to realize quantum East models).

Another natural extension to our work is to 
develop a deeper understanding of the physical meaning of $\GEC$.
On the one hand, this may require insights from a graph theory that includes diagonal matrix elements, such as the ones induced by diagonal detuning in our case.
Partially, such developments are already on their way in the investigation of graphs with self-loops (see, e.g.,~\cite{PreethaP2023}) and a treatment of self-loops with arbitrary weights would be interesting in the context of our work. 
On the other hand, the relative detuning of nodes in the Fock-space graph in the product basis with respect to their
diagonal Hamiltonian matrix elements
is conventionally described by perturbation theory.
Establishing the connection between diagonal detuning, $\GEC$, and the perturbative theory is left for future work.\\

Research data associated with this publication will be made available on Zenodo~\cite{this_zenodo}.

\begin{acknowledgments}
We acknowledge useful discussions with A. Lazarides, F. Pollmann, and P. Sollich.
This work was funded by the Deutsche Forschungsgemeinschaft (DFG, German Research Foundation) – 436382789, 493420525, 499180199; via FOR 5522 and large-equipment grants (GOEGrid cluster), and under Germany's Excellence Strategy -- EXC-2111 -- 390814868.

\end{acknowledgments}

\appendix

\section{Constructing the quantum East graph}
\label{sec:appendix_construct_many_body_graphs}

\begin{figure}[t]
    \centering
    \includegraphics[width=\linewidth]{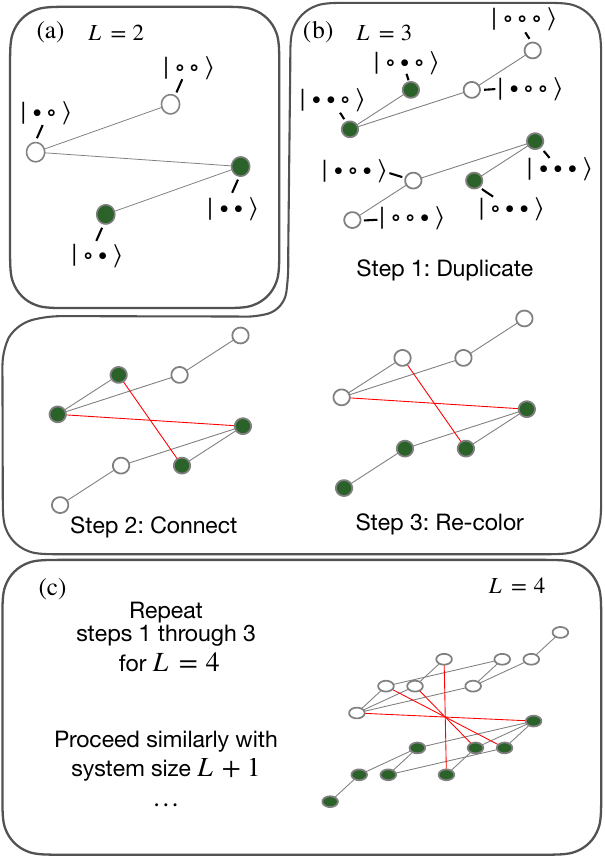}
    {\phantomsubcaption\label{fig:graph_construction_L=2}}
    {\phantomsubcaption\label{fig:graph_construction_L=3}}
    {\phantomsubcaption\label{fig:graph_construction_L=4}}
    \caption{
        Sketch illustrating the iterative construction of the QE graph.
        In \subref{fig:graph_construction_L=2}, we show the $L=2$ graph as discussed in the text and in \subref{fig:graph_construction_L=3}, we sketch the algorithm that is used to extend the graph from $L=2$ to $L=3$.
        \subref{fig:graph_construction_L=4} outlines the process for the next step and shows the result of the iterative step for $L=4$.
    }
    \label{fig:graph_construction}
\end{figure}

We here describe the construction of the many-body graph $g_\mathrm{MB}$ for the quantum East model in the computational basis, i.e., the joint
eigenstates of $n_\ell$ operators, and how to produce its graphical representation.
To do so, we introduce an iterative procedure of how to grow $g_\mathrm{MB}$ as we increase the system size of the many-body system.
Assuming that we already have a graph $g_\mathrm{MB}$ of a many-body lattice system with $L$ sites, each with a local Hilbert space dimension $d$, we notice that when we increase the system size by a single site, the number of states (and nodes on the graph) is multiplied by $d$.
Further, all of the previous transitions encoded in $g_\mathrm{MB}$ have to be preserved whatever the state is of the newly added site.
Therefore, in a first constructive step, we create $d$ copies of $g_\mathrm{MB}$,
keeping all internal edges of the graph intact.
Each of these copies corresponds to configurations in which the added site is in a different state.
Next, we need to connect the 
subgraphs with each other.
This is determined by 
 the structure of the off-diagonal operators of the Hamiltonian under consideration.
To construct the full graph $g_\mathrm{MB}$, we start from a system of a single site, choosing initial edges on the graph depending on the Hamiltonian and its boundary conditions, and then grow this graph successively, applying the previously described steps until we have reached the desired system size.

As an illustration of this construction, we discuss how to construct the graph $g_\mathrm{QE}$ of the QE model in the joint eigenbasis of all $n_\ell$ operators.
Starting from a system with a single site $L=1$ under the East boundary conditions discussed in \cref{sec:introduction_model}, the transition on the first site of the system is facilitated and therefore,
we start with a graph with two connected nodes 
\begin{align}
    \ket{\bullet} \leftrightharpoons \ket{\circ}\,.
    \end{align}
    To grow the system to $L=2$, we add a site to the right, and apply the steps described above, to obtain:
\begin{center}
\vspace{0.7em}
\begin{tikzcd}[row sep=large,column sep=large]
{|\bullet\circ\,\rangle}
    \arrow[thick,leftharpoonup,xshift=0.3ex,yshift=0.3ex,color=red]{dr}{} 
    \arrow[thick,leftharpoonup,yshift=0.5ex]{r}{} 
& 
{|\circ\circ\,\rangle}
    \arrow[thick,leftharpoonup]{l}{} 
\\
{|\circ\bullet\,\rangle}
    \arrow[thick,leftharpoonup,yshift=0.5ex]{r}{}
&  
{|\bullet\bullet\,\rangle}
    \arrow[thick,leftharpoonup,color=red]{ul}{} 
    \arrow[thick,leftharpoonup]{l}{} 
\end{tikzcd}
\end{center}
\vspace{0.7em}
The top and bottom rows of the diagram correspond to the duplicated subgraphs of the single-site system with their respective, preserved transitions shown in dark color, while the single, facilitated transition between the two subgraphs, derived from the off-diagonal operators in the QE model, is shown in red.
For visual compactness, we choose to flip the graphical representation of one of the copies.
Further, we notice that only the subgraph in the lower row of the diagram contains states which have an occupied last site.
Only these states will facilitate additional transitions  in the next step from $L=2$ to $L=3$.
To keep track of the corresponding nodes, we label them by color, as shown in Fig.~\ref{sec:appendix_gec_pdf}. The nodes keep their color through the duplication.

After duplication it is now easy to decide which nodes to connect on the graph:
The transitions between the two subgraphs are exactly those which connect two colored nodes that are copies of each other.
After adding all the necessary edges as described above, 
we remove the color labels and, instead, color the nodes of the subgraph that will trigger transitions in the next step, namely, the copy added in the last step, which corresponds to an occupied last site.
In \cref{fig:graph_construction} we have sketched the constructive process from $L=2$ to $L=3$ in detail, which can be directly extended to larger system sizes.

\section{Exact calculation of the \texorpdfstring{$\GEC$}{GEC} distributions}
\label{sec:appendix_gec_pdf}

The probability tables of the $\GEC$ can be computed exactly for most of the protocols considered in the paper, whose Hamiltonians can be generically written in the form $H(s,z)=-\frac{1}{2}e^{-s}A+\frac{1}{2}\Delta+z P$ shown in \cref{eq:genericH}. The term $A$ includes the off-diagonal part of $H$, whereas $\Delta$ and $P$ are diagonal.
For East boundary conditions,
\begin{align}
A&=\sum_{\ell=1}^{L-1}n_{\ell} \sigma^x_{\ell+1}+\sigma^x_{1}\,, \nonumber \\
\Delta&=\sum_{\ell=1}^{\changes{L-1}}n_{\ell}-\gamma e^{-s}n_L+c\openone\,,
\label{eq:A_Delta}
\end{align}
with $\gamma=\pm 1$, corresponding to the $\sigma^x$ eigenstate chosen for the non-dynamical $(L+1)$-th site [see \cref{eq:HQE_actual}] and $c$ can be chosen to make $H$ traceless 
as $c=\frac{\gamma e^{-s}-L}{2}-2^{-(L-1)}z \Tr P$.

As shown in \cref{sec:discussion}, the $\GEC$ for a configuration $\ket{i}$ can be written as
\begin{align}
    \label{eq:ge_simplified_appendix}
    \GEC(\ket{i}) &= \frac{2 \mel{i}{H^2}{i} - \mel{i}{H}{i}^2}{\Tr(H^2)}\,.
\end{align}
In order to evaluate this expression, we notice that only diagonal elements (of either $H^2$ or $H$) contribute. We can thus write
\begin{align}
\Tr H^2=&\frac{e^{-2s}}{4} \Tr
A^2+ \frac{1}{4} \Tr
\Delta^2+ z \Tr(P\Delta) + z^2 \Tr P^2\,,
\label{eq:denominatorGEC_depP}
\end{align}
and
\begin{align}
2\bra{i} H^2 \ket{i}+&\mel{i}{H}{i}^2=\frac{e^{-2s}}{2}\mel{i}{A^2}{i}
\nonumber \\
&+\left(\mel{i}{\frac{\Delta}{2}}{i}+z \mel{i}{P}
{i}\right)^2\,.
\label{eq:numeratorGEC_depP}
\end{align}
Using the explicit form of $A$ and $\Delta$ in \cref{eq:A_Delta},
\begin{align}
\bra{i}A^2\ket{i}=\sum_{\ell=1}^{L-1}\mel{i}{n_{\ell}}{i}+1\,,
\end{align}
since $n_{\ell}^2=n_{\ell}$, and the other terms in $A^2$ only have off-diagonal contributions.
Similarly,
\begin{align}
    \Tr A^2=&\Tr\left(\sum_{\ell=1}^{L-1} n_{\ell}+\openone\right)=2^{L-1}(L+1)\,,
\end{align}
and
\begin{align}
&\Tr\Delta^2=\Tr\left( \sum_{\ell=1}^{L-1}n_{\ell}+(1-\gamma e^{-s})n_L+c\openone\right)^2\nonumber \\
&=
2^{L-1}\left[e^{-2 s}-\gamma e^{-s}(L+2c+1)+\frac{L^2+(4c+1)L+4c^2}{2}\right]\,,
\end{align}
where we used that $\Tr(n_{\ell})=2^{L-1}$ and $\Tr (n_{\ell}n_{\ell'})=2^{L-2}(1+\delta_{\ell \ell'})$.

\subsection{\texorpdfstring{$\GEC$}{GEC} for the (undetuned) QE model}
\label{subsec:gec_qe}

Using the expressions above, we can write explicitly the $\GEC$ of configuration $\ket{i}$ for the QE Hamiltonian in a system of size $L$, as shown in \cref{eq:gec_qe},
\begin{align}
&\GEC_\mathrm{QE}(\ket{i}) =\GEC_\mathrm{QE}(M(i),m(i)) \\
& =
\frac{
    2 e^{-2s}(M(i)+1)+(M(i)+(1-\gamma e^{-s})m(i)+c)^2
}{
    4\Tr(H_\mathrm{QE}^2)
}\,,
\end{align}
and $\Tr(H_\mathrm{QE}^2) = 2^{L-3}[e^{-2s}(L+2)-\gamma e^{-s}(L+2c + 1)+\frac{1}{2}(L^2+(4c+1)L+4c^2)]$.
We have also defined
\begin{align}
M(i)&\equiv \sum_{\ell=1}^{L-1}\bra{i}n_{\ell}\ket{i}\,, \nonumber \\
m(i)&\equiv \bra{i}n_L\ket{i}\,.
\label{eq:def_M_m}
\end{align}
Therefore, the distribution of $\GEC$ values in this case is determined by the distribution of variables $M$ and $m$ over the $2^L$ configurations. Since these variables can take values
$M\in\{0,\, 1,\ldots, L-1\}$ and $m\in\{0,\,1\}$, the $\GEC$ can adopt a maximum of $2L$ distinct values. Each value $\GEC_\mathrm{QE}(M,m)$ will occur with probability $p(M,m)$
\begin{align}
p(M,m)
=p(M)p(m)
=\frac{1}{2^{L}} \binom{L-1}{M}\,.
\end{align}

Similarly, we can obtain the probability tables for each of the detuning protocols by explicitly computing the $P$-dependent terms in \cref{eq:numeratorGEC_depP} and \cref{eq:denominatorGEC_depP}.

\subsection{\texorpdfstring{$\GEC$}{GEC} for local detuning}
\label{subsec:gec_local}

For the local detuning
$P_\mathrm{loc}=n_{L/2}$ we get
\begin{align}
\Tr P_\mathrm{loc}^2=&\Tr P_\mathrm{loc} =2^{L-1}\,, \nonumber \\
\Tr (P_\mathrm{loc}\Delta) =&
=2^{L-2}(L+2c+1-\gamma e^{-s})\,,
\end{align}
so that the denominator of the $\GEC$ is
\begin{align}
\Tr H_\mathrm{loc}^2=
\Tr H_\mathrm{QE}^2 +2^{L-2}\left[z(L+2c+1-\gamma e^{-s})+2z^2\right]\,,
\end{align}
where we called
\begin{align}
    \Tr H_\mathrm{QE}^2=\frac{e^{-2s}}{4} \Tr
A^2+ \frac{1}{4} \Tr
\Delta^2\,.
\end{align}
For the numerator we only need to 
substitute $\bra{i}P_\mathrm{loc}
\ket{i}$ in \cref{eq:numeratorGEC_depP}, and, defining 
\begin{align}
k(i) & \equiv \bra{i}n_{L/2}\ket{i} \,, \quad k(i) \in \{0,1 \}\,,\nonumber\\
m(i) & \equiv \bra{i}n_L\ket{i}\,, \quad m(i) \in \{0,1 \}\,,\nonumber\\
X(i) & \equiv \sum_{ \ell\neq L/2,L}\bra{i}n_{\ell}\ket{i},\quad X(i) \in \{ 0,\,1,\ldots, L-2\}\,,
\end{align}
we can finally write the $\GEC$ as a function of these three variables, taking up to $4(L-1)$ distinct values,
\begin{align}
&\GEC_\mathrm{loc}(X,k,m)= \\
&\frac{2 e^{-2s}(X+k+1)+\left(X+(1+2z)k+(1-\gamma e^{-s})m+c\right)^2}{4 \Tr H_\mathrm{loc}^2}\,, \nonumber
\end{align}
which occur with probability
\begin{equation}
    p(X,k,m)=\frac{1}{2^L} \binom{L-2}{X}\,.
\end{equation}

\subsection{\texorpdfstring{$\GEC$}{GEC} for filling detuning}
\label{subsec:gec_filling}

In this protocol, the detuning affects nodes with a fixed number of particles $N_0(i)=\sum_\ell \mel{i}{n_\ell}{i} = L/2$, and we can write
\begin{align}
P_\mathrm{filling}=\sum_{i;N_0(i)=L/2}\ket{i}\bra{i}\,,
\end{align}
so that
\begin{align}
\Tr P_\mathrm{filling}^2=
&\Tr P_\mathrm{filling} =\binom{L}{L/2}\,, \nonumber \\
\Tr (P_\mathrm{filling}\Delta) 
=& \left(\frac{L}{2}+c\right)\Tr P_\mathrm{filling}-\gamma e^{-s} \binom{L-1}{L/2-1}\,,
\end{align}
the last term originating from the trace of the product of projectors $n_L $ and $P_\mathrm{filling}$.
Thus
\begin{align}
\Tr H_\mathrm{filling}^2=& \Tr H_\mathrm{QE}^2
+z^2\binom{L}{L/2} \\
&+z\left[ \left(\frac{L}{2}+c\right)\binom{L}{L/2}-\gamma e^{-s}\binom{L-1}{L/2-1} \right ] \nonumber 
\,.
\end{align}
Using the variables defined in \cref{eq:def_M_m},
we can finally write the possible values of the $\GEC$ as
\begin{align}
\GEC&_\mathrm{filling}(M,m)= \\
=&\frac{2 e^{-2s}(M+1)+\left(M+(1-\gamma e^{-s})m+c+2 z \delta_{M+m,L/2}\right)^2}{4 \Tr H_\mathrm{filling}^2}\,, \nonumber
\end{align}
with probabilities
\begin{equation}
p(M,m)=\frac{1}{2^L}\binom{L-1}{M}\,.
\end{equation}

\subsection{\texorpdfstring{$\GEC$}{GEC} for interaction detuning}
\label{subsec:gec_interactions}

In this case, the detuning term reads
\begin{align}
P_\mathrm{interact}=\sum_{\ell=1}^{L-1}n_{\ell}n_{\ell+1}\,,
\end{align}
and 
$\Tr P_\mathrm{interact}=2^{L-2}(L-1)$\,.

Expanding $P_\mathrm{interact}^2$ in terms with two, three and four different $n_{\ell}$ operators, we find
\begin{align}
&\Tr P_\mathrm{interact}^2=2^{L-4}\left[ (L+1)(L+2)-8\right]\,,
\end{align}
and 
\begin{align}
\Tr &(P_\mathrm{interact}\Delta)
= \\
&=2^{L-3} 
\left[(L-2)(L+1)
    +(2-\gamma e^{-s})L
    +2c(L-1) 
\right]\,.
\nonumber
\end{align}

In this case, $\bra{i}P_\mathrm{interact}\ket{i}$ counts the number of pairs of neighboring occupied sites in configuration $\ket{i}$.
Therefore, we define three variables,
\begin{align}
N & \equiv \sum_{\ell=1}^{L}\bra{i}n_{\ell}\ket{i},\quad N \in \{ 0,\,1,\ldots L\}\,,\nonumber\\
m & \equiv \bra{i}n_L\ket{i}\,, \quad m \in \{0,1 \}, \nonumber\\
p & \equiv \bra{i}P_\mathrm{interact}\ket{i}\,, \quad p \in \{0,L-1 \}\,.
\end{align}

The $\GEC$ values can then be expressed as
\begin{align}
\GEC&_\mathrm{interact}(N,m,p) = \\
&\frac{2 e^{-2s}(N-m+1)+(N-\gamma e^{-s} m+c+2zp)^2}{4\Tr(H_\mathrm{interact}^2)}\,. \nonumber
\end{align}
In this case, the values of $N$, $m$ and $p$ are not independent from each other, and the probability of value $\GEC_\mathrm{interact}(N,m,p)$ will be given by the joint probability $p(N,m,p)$.
The latter can be computed efficiently by computing the trace of the product of projectors onto total number of particles equal $N$, number of pairs of neighboring particles equal to $p$ and number of particles on site $L$ equal to $m$.

\changes{

\section{\texorpdfstring{Details for the $P_\mathrm{filling}$ protocol}{Details for the filling protocol}}
\label{sec:appendix_filling}

On the Fock graph in the joint eigenbasis of the $n_\ell$ operators, the $P_\mathrm{filling}$ protocol introduces an energy barrier between the basis states of low filling $N_0 < L/2$ and the basis states at high filling $N_0 \ge L/2$.
In \cref{fig:filling_detuning_subsystem}, we can observe that the eigenstates in the QE system subject to the $P_\mathrm{filling}$ detuning protocol generally have support only in one of the two subsystems, with only the states at around half-filling having support in both subsystems.
Furthermore, the inset shows that energy density and filling of the respective eigenstates are highly correlated.
Therefore eigenstates with support in the low-filling subgroup generally correspond to $\varepsilon < 0.5$ and eigenstates with support in the high-filling subgroup correspond to $\varepsilon > 0.5$.

As we have observed in \cref{fig:EEs_protocols_spectrum}, the entanglement entropy curve of the $P_\mathrm{filling}$ protocol features two maxima to the left and right of $\epsilon = 0.5$.
As eigenstates at these energy densities only have support in one of the low- or high-filling subgroups on the graph, these entropy maxima indicate that the states within these subgroups have a tendency to be extended where they have support, namely within their respective subgroup.
In any case the states are more extended than in the $P=0$ case.

\begin{figure}[ht]
    \centering
    \includegraphics[width=\linewidth]{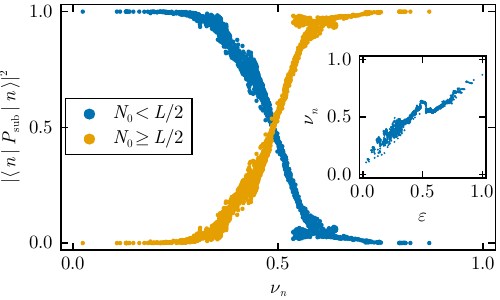}
    \caption{
        \changes{
        Weight of the eigenstate $\ket{n}$ in the subsystems $N_0 < L/2$ and $N_0 \ge L/2$ in the $P_\mathrm{filling}$ protocol ($L=12$, $s=0.5$) against the filling of the eigenstates ${\nu_n = \langle n| \sum_ {\ell =1}^L n_\ell|  n \rangle/L }$. 
        $P_\mathrm{sub} $ is the operator that projects onto the selected subsystem.
        The inset shows how the filling of the eigenstates is correlated with the energy density $\varepsilon$ of a given eigenstate.
        }
    } 
    \label{fig:filling_detuning_subsystem}
\end{figure}

}

\bibliography{main.bib}

\end{document}